\documentclass[12pt]{article}

\pdfoutput=1

\usepackage{graphicx}
\usepackage{epsfig}

\usepackage{amsfonts}
\usepackage{amssymb}
\usepackage{amsmath}
\usepackage{dsfont}
\usepackage{pifont}
\usepackage{bbm}
\usepackage{multirow}
\usepackage{latexsym}
\usepackage{verbatim}
\usepackage{mcite}
\usepackage{cite}
\usepackage{units}
\usepackage[all]{xy}
\usepackage{arydshln}
\usepackage{cancel}
\usepackage{slashed}

\def\beq{\begin{equation}}
\def\eeq{\end{equation}}
\def\beqa{\begin{eqnarray}}
\def\eeqa{\end{eqnarray}}

\def\m{{\mu}}
\def\n{{\nu}}

\def\bfone{\relax{\rm 1\kern-.35em 1}}

\newcommand{\cN}{{\cal N}}

\newcommand{\cV}{{\cal V}}

\newcommand{\be}{\begin{equation}}
\newcommand{\ee}{\end{equation}}
\newcommand{\ben}{\begin{displaymath}}
\newcommand{\een}{\end{displaymath}}
\newcommand{\bea}{\begin{eqnarray}}
\newcommand{\eea}{\end{eqnarray}}

\newcommand{\bean}{\begin{eqnarray*}}
\newcommand{\eean}{\end{eqnarray*}}

\DeclareMathAlphabet{\mathpzc}{OT1}{pzc}{m}{it}

\begin{document}
\pagestyle{plain}

%----------------------------------------------------------------------%
%  numbering sections, equations, footnotes, etc...
%----------------------------------------------------------------------%

\makeatletter \@addtoreset{equation}{section} \makeatother
\renewcommand{\thesection}{\arabic{section}}
\renewcommand{\theequation}{\thesection.\arabic{equation}}
\renewcommand{\thefootnote}{\arabic{footnote}}

%----------------------------------------------------------------------%
%  Resetting of counters
%----------------------------------------------------------------------%

\setcounter{page}{1} \setcounter{footnote}{0}

%----------------------------------------------------------------------%
%  title page
%----------------------------------------------------------------------%

\begin{titlepage}
\begin{flushright}
\small ~~
\end{flushright}

\bigskip

\begin{center}

\vskip 0cm

{\LARGE \bf On new maximal supergravity\\[4mm] and its BPS domain-walls} \\[6mm]

\vskip 0.5cm

{\bf Adolfo Guarino }\\

\vskip 25pt

{\em 
Albert Einstein Center for Fundamental Physics\\
Institute for Theoretical Physics, Bern University\\
 Sidlerstrasse 5, CH–3012 Bern, Switzerland\\
{\small {\tt guarino@itp.unibe.ch}}}

\vskip 0.8cm

\end{center}

\vskip 1cm

\begin{center}

{\bf ABSTRACT}\\[3ex]

\begin{minipage}{13cm}
\small

We revise the SU(3)-invariant sector of $\,{\mathcal{N}=8}\,$ supergravity with \textit{dyonic} $\,\textrm{SO}(8)\,$ gaugings. By using the embedding tensor formalism, analytic expressions for the scalar potential, superpotential(s) and fermion mass terms are obtained as a function of the electromagnetic phase $\omega$ and the scalars in the theory. Equipped with these results, we explore non-supersymmetric AdS critical points at $\omega \neq 0$ for which perturbative stability could not be \mbox{analysed} before. The $\omega$-dependent superpotential is then used to derive first-order flow equations and obtain new BPS domain-wall solutions at $\omega \neq 0$. We numerically look at steepest-descent paths motivated by the (conjectured) RG flows.

% connecting $\,{\mathcal{N}=8}\,\&\,\textrm{SO}(8)$, ${\mathcal{N}=2}\,\&\,\textrm{U}(3)$, ${\mathcal{N}=1}\,\&\,\textrm{G}_{2}$ and ${\mathcal{N}=1}\,\&\,\textrm{SU}(3)$ critical points 

\end{minipage}

\end{center}

\vfill

\end{titlepage}

%%%%%%%%%%%%%%%%%%%%%%%%%%%%%%%%%%%%%%%%%%%%%%%%%%%%%%%%%
%%
%%               Contents
%%
%%%%%%%%%%%%%%%%%%%%%%%%%%%%%%%%%%%%%%%%%%%%%%%%%%%%%%%%%

\tableofcontents

\section{Motivation and outlook}
\label{sec:Motivation and outlook}

Dimensional reduction of higher-dimensional field theories down to four dimensions (4d) has proven a very successful road towards the unification of gravitational and Yang-Mills interactions \cite{Kaluza:1921,Klein:1926tv}. The first modern constructions go back to the seminal papers \cite{Brink:1976bc} and \cite{Scherk:1979zr} in the 70's where the dimensional reductions of super Yang-Mills theory (SYM) and supergravity (SUGRA) were discussed. Both cases, even though fundamentally different in what concerns the theories to be reduced, display some universal features: $i)$ appearance of a scalar potential $V$ $\,\,ii)$ generation of fermion mass terms  $\,\,iii)$  modification of the supersymmetry transformation rules for the fermions in the theory. The reduced Lagrangian can schematically be viewed as
\beq
\begin{array}{ccc}
\label{L_general_form}
\mathcal{L}_{\textrm{higher-dim}} &\longrightarrow & \mathcal{L}_{\textrm{4d}} = \mathcal{L}_{\textrm{kinetic}} + \mathcal{L}_{\textrm{fermi}} - V \ ,
\end{array}
\eeq
where the concrete expressions for the fermi mass terms in $\mathcal{L}_{\textrm{fermi}}$ and the scalar potential $V$ depend on the theory which is to be reduced.

\vspace{3mm}

\noindent\textit{Fetching ideas from  SYM}

\vspace{3mm}

Reducing $\mathcal{N}=1$, $d=10$ SYM on a six-torus produces $\mathcal{N}=4$, $d=4$ SYM. The reduced Lagrangian is of the form\footnote{We are not being precise in the definition of $\psi_{i}$ in the r.h.s of (\ref{L10toL4}). This would imply introducing chirality projectors as well as a charge conjugation matrix which are not relevant at the level of the discussion here.} \cite{Brink:1976bc}
\beq
\begin{array}{cclcccl}
\label{L10toL4}
\mathcal{L}_{\textrm{10d}} &=&   -\frac{1}{4} F^2 + i \bar{\psi}  \, {\not} D \psi  &  \longrightarrow & \mathcal{L}_{\textrm{4d}} &=& -\tfrac{1}{4} F^2 + i \, \bar{\psi}_{i}  \, {\not} D \psi^{i} + \tfrac{1}{2} (D\phi_{ij})^2 \\[1mm] 
& & & & & & + \,\frac{i}{2} g \, ( f\, \phi^{ij}\bar{\psi}_{i} \psi_{j} - c.c)   \\[2mm] 
& & & & & & -\, \tfrac{1}{4} g^2 \, (f \, \phi_{ij} \phi_{kl})^2   \ ,
\end{array}
\eeq
where $g$ is the gauge coupling constant, $f$ represents the structure constants of the gauge group and $i=1,..,4$ is a fundamental index of the R-symmetry group $\textrm{SU}(4)\sim \textrm{SO}(6)$ emerging from the reduction. The four Weyl fermions $\psi^{i}$ descend from the original Majorana-Weyl fermion $\psi$ in ten dimensions, whereas the scalar fields $\phi_{ij}=\phi_{[ij]}$ correspond to the six internal components of the 10d gauge fields and are subject to the reality condition
\beq
\label{reality SYM}
\phi_{ij} \,=\,\tfrac{1}{2} \,\epsilon_{ijkl} \,\phi^{kl} 
\hspace{15mm}
\textrm{with} 
\hspace{10mm}
\phi^{kl}=(\phi_{kl})^{*}
\ .
\eeq
A key observation \cite{Brink:1976bc}  is that the reality condition (\ref{reality SYM}) prevents the R-symmetry group of the reduced theory to be extended to $\textrm{U}(4)=\textrm{U}(1) \times \textrm{SU}(4)$. The additional U(1) is simply not compatible with this condition.

The interaction in the reduced theory stems from the non-abelian structure of the theory in higher-dimensions ($gf \neq 0$). The $\mathcal{L}_{\textrm{4d}}$ in the r.h.s of (\ref{L10toL4})  matches the general form (\ref{L_general_form}) for dimensionally reduced theories: the first line contains the kinetic terms for the different fields, the second one corresponds to (scalar dependent) fermi mass terms which are of order $g f$, and the last line is identified with a scalar potential of order $(gf)^2$. Last but not least, the supersymmetry (SUSY) transformation for the fermions in the reduced theory reads
\beq
\label{SUSY-SYM}
\delta_{\epsilon}\psi^{i} =  F^{\mu\nu} \gamma_{\mu \nu} \epsilon^{i} -   {\not} D \phi^{ij} \epsilon_{j}  + \tfrac{1}{2} g f \phi^{ij} \phi_{jk} \epsilon^{k} \ ,
\eeq
so the last term implies a modification (linear order in $gf$ like the fermi mass terms) with respect to the standard transformation rule.

\vspace{3mm}

\noindent\textit{The SUGRA side of the story}

\vspace{3mm}

Kaluza-Klein reductions of $\mathcal{N}=2$, $d=10$ supergravity on a six-torus and of 11d supergravity on a seven-torus produces $\mathcal{N}=8$ (maximal) \textit{ungauged} supergravity in four dimensions \cite{Cremmer:1979up,Cremmer:1997ct}. The resulting theory possesses an abelian $G=\textrm{U}(1)^{28}$ gauge symmetry under which all the scalars coming from the reduction of the higher-dimensional fields are neutral. As a consequence, no scalar potential or fermion mass terms are generated
\beq
\begin{array}{ccc}
\label{L_ungauged}
\mathcal{L}_{\textrm{10d/11d}} &\overset{ungauged}{\longrightarrow}  & \mathcal{L}_{\textrm{4d}} = \mathcal{L}_{\textrm{kinetic}}  \ .
\end{array}
\eeq

However, certain background fluxes for the higher-dimensional fields 
%-- as well as more exotic non-geometric/generalised flux backgrounds without a clear higher-dimensional origin\footnote{Some of them have been recently related to asymmetric orbifold constructions in ref.~\cite{Condeescu:2013yma}.} -- %
can be turned on in a way still compatible with $\mathcal{N}=8$ supersymmetry \cite{deWit:2003hq,deWit:2007mt,Samtleben:2008pe,Aldazabal:2011yz,Dibitetto:2012ia} inducing what is called a \textit{gauging}. The result of the reduction is then a \textit{gauged} supergravity with a non-abelian gauge group $G$ and coupling constant $g$. The background fluxes are identified with the structure constants of $G$ and the situation becomes similar to the SYM reduction (\ref{L10toL4}) with $f\equiv\textrm{fluxes}$.  As a consequence of the gauging, the scalars in the theory become charged under $G$ and both a scalar potential (quadratic on $g \, \cdot$ fluxes) and fermi mass terms (linear on $g \, \cdot$ fluxes) are generated 
\beq
\begin{array}{ccc}
\label{L_general_gauged}
\mathcal{L}_{\textrm{10d/11d}} &\overset{gauging}{\longrightarrow} & \mathcal{L}_{\textrm{4d}} = \mathcal{L}_{\textrm{kinetic}} + \mathcal{L}_{\textrm{fermi}} - V \ .
\end{array}
\eeq
The supersymmetry transformation rules for the gravitini and the dilatini get also modified in a similar fashion to (\ref{SUSY-SYM}). In analogy to the SYM condition (\ref{reality SYM}), the scalar fields in maximal supergravity can be arranged into a tensor $\phi_{IJKL}=\phi_{[IJKL]}$ subject to the reality condition
\beq
\label{reality SUGRA}
\phi_{IJKL} \,=\,\tfrac{1}{4!} \,\epsilon_{IJKLMNPQ} \,\phi^{MNPQ} 
\hspace{15mm}
\textrm{with} 
\hspace{10mm}
\phi^{MNPQ}=(\phi_{MNPQ})^{*}
\ .
\eeq
This time the index $I=1,...,8$ refers to the fundamental representation of the $\textrm{SU}(8)$ \mbox{R-symmetry} group emerging from the reduction. As for SYM, the reality condition (\ref{reality SUGRA}) prevents the R-symmetry group to be extended to $\textrm{U}(8)=\textrm{U}(1) \times \textrm{SU}(8)$. 

The lack of the U(1) factor both in $\mathcal{N}=4$ SYM and $\mathcal{N}=8$ SUGRA in four dimensions relates to the fact that these are CPT-self-conjugate multiplets of the supersymmetry algebra\footnote{We thank J.P.~Derendinger for pointing us this out and for useful discussions and explanations on this issue. For further reading, see the lecture notes \cite{Derendinger:1990tj} and references therein.}. They satisfy the condition $\lambda_{MAX} = \mathcal{N}/4$ with $\lambda_{MAX}$ being the maximum helicity state inside the supermultiplet. Because of self-conjugacy, CPT doubling is not necessary. Then the scalars sit in a real representation of the R-symmetry group and the reality conditions (\ref{reality SYM}) and (\ref{reality SUGRA}) preventing the U(1) factor have to be imposed.

\vspace{3mm}

\noindent\textit{A novel U(1) in maximal supergravity}

\vspace{3mm}

\begin{figure}[t!]
\begin{center}
\includegraphics[width=90mm]{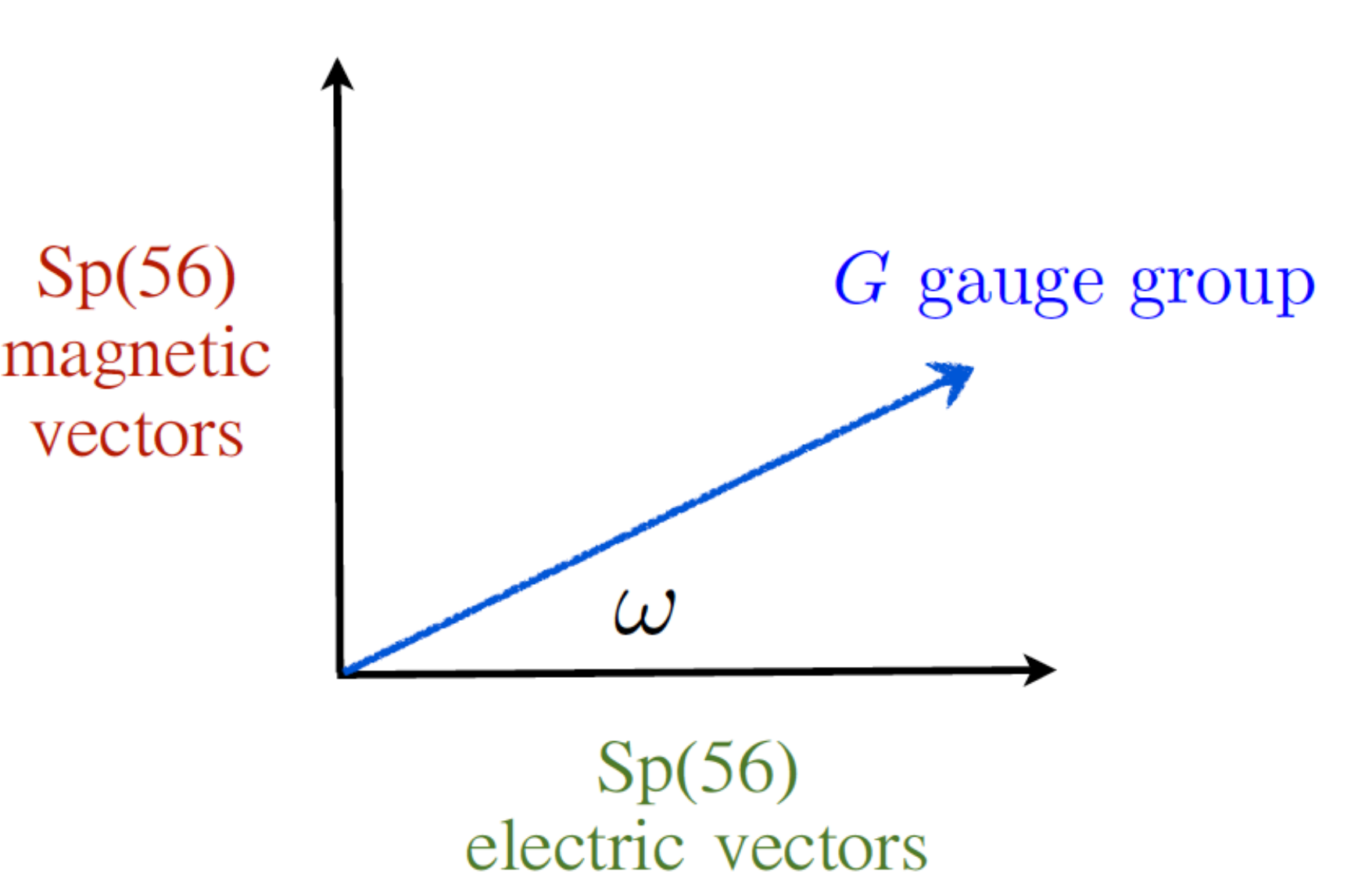}
\caption{{\it Orientation of the gauge group $G$ inside the Sp(56) electromagnetic group of maximal supergravity in four dimension. Setting $\omega=0$ corresponds to a purely electric gauging whereas $\omega=\frac{\pi}{2}$ corresponds to a purely magnetic choice.}}
\label{Fig:orientation} 
\end{center}
\end{figure}

The existence of a relevant U(1) in maximal supergravity lying outside the SU(8) R-symmetry group but still inside the Sp(56) electromagnetic group of the theory, was exploited in ref.~\cite{Dall'Agata:2012bb} to build a one-parameter family of gauged supergravities with $G=\textrm{SO}(8)$ gauging. This parameter was identified with an electromagnetic  phase $\omega$ which specifies the linear combination of electric ($28$ of them) and magnetic ($28$ of them) vectors entering the gauging $G$ (see Figure~\ref{Fig:orientation}), namely
\beq
A^{G}_{\mu}=\cos \omega \,  A^{elec}_{\mu} + \sin \omega \, A^{magn}_{\mu} \ .
\eeq
The $G_{2}$-invariant sector of this new family of SO$(8)$ gauged supergravities was analysed in refs~\cite{Dall'Agata:2012bb,Borghese:2012qm} and found to contain genuinely \textit{dyonic} critical points at $\omega \neq 0$ with no counterpart in the standard electric case\footnote{The $\omega$-dependent family of maximal supergravities with a different $G=\textrm{SO}(4,4)$ gauging was also investigated in ref.~\cite{Dall'Agata:2012sx} and found to contain SO(4)-invariant unstable de Sitter critical points with arbitrary light tachyons controlled by $\omega$. This intriguing phenomenon of tachyon amelioration was connected to a jump of gaugings involving an AdS/Mkw/dS transition in ref.~\cite{Borghese:2013dja}.} of $\omega=0$ \cite{Cremmer:1979up,deWit:1982ig,Warner:1983vz}. Similar results followed from the analysis of the SU($3$)-invariant sector in ref.~\cite{Borghese:2012zs} based on a \textit{conjectured} $\omega$-dependent superpotential compatible with the $\mathcal{N}=2$ structure of the truncated theory \cite{Ahn:2000mf,Ahn:2009as,Bobev:2010ib} as well as with \mbox{$\,\omega \rightarrow -\omega\,$} and \mbox{$\,\omega \rightarrow \omega + \frac{\pi}{4}\,$} identifications of the electromagnetic phase \cite{Dall'Agata:2012bb,Borghese:2012zs,Borghese:2013dja,deWit:2013ija}. In this way, an $\omega$-dependent superpotential could be envisaged (up to an overall phase) and the structure of SU(3)-invariant critical points investigated. 

However, a supergravity derivation of the $\omega$-dependent $\mathcal{L}_{\textrm{4d}}$  including fermi mass terms $\mathcal{L}_{\textrm{fermi}}$ and scalar potential $V$ for the SU(3) truncation, as done in refs~\cite{Ahn:2000mf,Ahn:2009as} for the $\omega=0$ case, remains to be done. As we will see later, the precise knowledge of $\mathcal{L}_{\textrm{fermi}}$ happens to be crucial for computing \textit{full} mass spectra at $\omega\neq0$ and will allow us to check the stability of critical points of $V$ which could not be analysed in ref.~\cite{Borghese:2012zs}. The derivation of $\mathcal{L}_{\textrm{4d}}$ can be carried out within the framework of the embedding tensor \cite{Nicolai:2000sc,Nicolai:2001sv,deWit:2007mt} and we will present it here. This is a purely four-dimensional supergravity formalism, so inverting the arrow in (\ref{L_general_gauged}) might not necessarily be possible. In other words, the connection to reductions of a higher-dimensional theory is lost. Nevertheless, the knowledge of $\mathcal{L}_{\textrm{fermi}}$ (more concretely of the $T$-tensor to be introduced later) as a function of $\omega$ and the scalar fields in the truncated theory, might help in finding new reduction Ans\"atze for a higher-dimensional origin of the electromagnetic phase along the lines of refs~\cite{deWit:1986iy,Nicolai:2011cy,Godazgar:2013nma}.

\vspace{3mm}

\noindent\textit{BPS domain-walls and flows equations}

\vspace{3mm}

A second interest in deriving the explicit form of the $\omega$-dependent $\mathcal{L}_{\textrm{4d}}$ is in the light of the AdS/CFT correspondence \cite{Maldacena:1997re,Witten:1998qj,Aharony:1999ti}. In its purely electric version ($\omega=0$), the SU(3)-invariant sector of the SO(8) gauged supergravity has played a central role in constructing RG flows dual to domain-wall solutions that interpolate between two AdS critical points of the scalar potential\footnote{We refer the reader to refs~\cite{Boonstra:1998mp,Skenderis:1999mm} as well as section~$9$ in ref.~\cite{D'Hoker:2002aw} for general reading and also refs~\cite{Ahn:2000aq,Ahn:2000mf,Ahn:2001kw,Bobev:2009ms,Ahn:2009as} for domain-walls in the SU(3)-invariant sector of the electric SO(8) supergravity. Other domain-walls were explored in refs~\cite{Ahn:2001by,Ahn:2002qga} for different (electric) gaugings.}. If the AdS points preserve some amount of supersymmetry the domain-wall is called BPS \cite{Skenderis:1999mm}. Then it can be constructed by solving a set of first-order flow-equations defined in terms of a superpotential $W(\phi^{i})$ with asymptotic behaviour $\left. \partial_{\phi^{i}} W \right|_{z \rightarrow \pm \infty} = 0$, where $z$ is the coordinate along the direction transverse to the domain-wall. The flow-equations for the set $\{\phi^{i}(z)\}$ of scalar fields are schematically given by \cite{Bobev:2009ms}
\beq
\frac{\partial \phi^{i}}{\partial z} \propto K^{ij}\frac{\partial W}{\partial \phi^{j}} 
\hspace{10mm} \textrm{and} \hspace{10mm} 
\frac{\partial A}{\partial z} \propto W(\phi^{i}) \ ,
\eeq
where $A(z)$ is the scale factor in the domain-wall metric Ansatz and $K^{ij}$ is the (inverse) K\"ahler metric accounting for non-canonically normalised kinetic terms for the scalars. In ref.~\cite{Ahn:2000mf}, the exact form of the superpotential $W$ was extracted from the fermi mass terms $\mathcal{L}_{\textrm{fermi}}$ in the case of a purely electric SO(8) gauging. Using that $\omega=0$ superpotential, various BPS domain-walls were constructed in the literature \cite{Ahn:2000aq,Ahn:2000mf,Ahn:2001kw,Bobev:2009ms,Ahn:2009as}. In the dual field theory picture they correspond to three-dimensional RG flows connecting an UV fixed point at $z \rightarrow \infty$ to an IR fixed point at $z \rightarrow -\infty$ (the scalars flow to a constant value for BPS domain-walls). Furthermore, higher-dimensional embeddings as reductions of 11d supergravity on $AdS_{4} \times S^7$ (with a round, squashed, stretched or warped seven-sphere) with a 4-form flux were found and connected to the theory of multiple M$2$-branes \cite{Corrado:2001nv,Bagger:2006sk,Bagger:2007jr,Gustavsson:2007vu,Aharony:2008ug,Ahn:2008gda,Ahn:2008ya,Bobev:2009ms,Ahn:2009jf}. 

\vspace{3mm}

The outline of the paper is as follows. Section~\ref{sec:Crash_course} collects standard results in ${\mathcal{N}=8}$ gauged supergravity in the framework of the embedding tensor which will be extensively used in this work (readers being familiar with the subject may go directly to section~\ref{sec:Dyonic_generalities}). In section~\ref{sec:N=2_trunc} we carry out a supergravity derivation of the scalar Lagrangian, the superpotential(s) and the fermion mass terms for the $\textrm{SU}(3)$-invariant sector of $\mathcal{N}=8$ supergravity with a \textit{dyonic} $\textrm{SO}(8)$ gauging. In section~\ref{sec:scalar_dynamics_BPS} we make use of these results to study the stability of non-supersymmetric AdS critical points and to obtain BPS domain-walls at $\omega\neq 0$. We then discuss the results and make some final remarks. More technical computations and lengthy expressions are put into the appendices.

\section{Crash introduction to maximal supergravity}
\label{sec:Crash_course}

After the previous discussion on R-symmetries and U(1)'s, we now summarise general results on $\,\mathcal{N}=8\,$ gauged supergravity in four dimensions (mostly from refs~\cite{deWit:2007mt,LeDiffon:2011wt}) and elaborate more on the idea of \textit{dyonic} gaugings \cite{DallAgata:2011aa}.

\subsection{Gaugings and scalar potential}
\label{sec:Generalities}

The bosonic field content of the supergravity multiplet in maximal supergravity consists of the metric $g_{\m\n}$, $56$ vector fields $A^{\mathbb{M}}_{\m}$ ($28$ electric and $28$ magnetic) and $70$ complex scalars $\Sigma_{IJKL}$ satisfying the self-duality  condition
\be
\label{SD_condition}
\Sigma_{IJKL} = \frac{1}{4!} \, \epsilon_{IJKLMNPQ} \, \Sigma^{MNPQ} \ ,
\ee
with $\Sigma^{IJKL}=(\Sigma_{IJKL})^{*}$, and which serve as coordinates in a coset space $\textrm{E}_{7(7)}/\textrm{SU}(8)$. The index $\mathbb{M}=1,...,56$ will refer to the fundamental representation of the (global) duality group $\textrm{E}_{7(7)}$, whereas $I=1,..,8$ to that of the (local) R-symmetry group $\textrm{SU}(8)$.

In the \textit{ungauged} (non-interacting) case, the choice of an $\textrm{Sp}(56,\mathbb{R})$ frame for the vector fields will not affect Physics. Thus, as an example of symplectic transformation, electric-magnetic duality will leave any observable invariant. However this picture changes dramatically once a set of charges  ${X_{\mathbb{MN}}}^{\mathbb{P}}$ is turned on and the theory becomes \textit{gauged} (interacting). In this case, symplectic transformations are no longer symmetries and these get  reduced to the duality group $\textrm{E}_{7(7)} \subset \textrm{Sp}(56,\mathbb{R})$. 

The first sign that the theory has been gauged is that part of the duality group has been promoted to a non-abelian gauge theory $G\subset \textrm{E}_{7(7)}$. The ordinary derivative is then replaced by a covariant $\,D_{\mu} = \partial_{\mu} - g A_{\mu}^{\mathbb{M}} {\Theta_{\mathbb{M}}}^{\textsf{A}} t_{\mathsf{A}}\,$ involving the vector fields $A_{\mu}^{\mathbb{M}}$, the $\textrm{E}_{7(7)}$ generators $t_{\textsf{A}}$ with $\textsf{A}=1,...,133$ and the so-called embedding tensor ${\Theta_{\mathbb{M}}}^{\textsf{A}}$. This tensor acts as a selector of $\textrm{E}_{7(7)}$ generators to be promoted to local symmetries and hence to be associated to gauge bosons. After a contraction with the $\textrm{E}_{7(7)}$ generators in the fundamental representation, one obtains the charges ${X_{\mathbb{MN}}}^{\mathbb{P}}={\Theta_{\mathbb{M}}}^{\textsf{A}} \, {[t_{\mathsf{A}}]_{\mathbb{N}}}^{\mathbb{P}}$ which play the role of structure constants of the gauge algebra
\be
\label{gauge_brackets}
[X_{\mathbb{M}} , X_{\mathbb{N}}] = - {X_{\mathbb{MN}}}^{\mathbb{P}} \,\, X_{\mathbb{P}} \ ,
\ee
spanned by the generators $X_{\mathbb{M}}$. Maximal supersymmetry and gauge invariance impose a set of respectively linear (LC) and quadratic constraints (QC) on the charges ${X_{\mathbb{MN}}}^{\mathbb{P}}$. The former  are related to the restriction of ${\Theta_{\mathbb{M}}}^{\textsf{A}}$ to live in the $\bold{912}$ irrep of $\textrm{E}_{7(7)}$. The latter come from the closure of the gauge brackets in (\ref{gauge_brackets}) and read
\be
\label{QC}
\Omega^{\mathbb{MN}} \, {\Theta_{\mathbb{M}}}^{\textsf{A}} \, {\Theta_{\mathbb{N}}}^{\textsf{B}} = 0 \ ,
\ee
where $\Omega_{\mathbb{MN}}$ is the $\textrm{Sp}(56,\mathbb{R})$ invariant matrix (skew-symmetric) satisfying $\Omega_{\mathbb{MP}} \, \Omega^{\mathbb{NP}} = \delta_{\mathbb{M}}^{\mathbb{N}}$. These two sets of constraints guarantee the consistency of the gauged supergravity.

The second sign is that a non-trivial scalar potential $V(\Sigma)$ is generated for the scalars spanning the $\textrm{E}_{7(7)}/\textrm{SU}(8)$ coset space. These scalars can be encoded inside a \textit{mixed} coset representative ${\mathcal{V}_{\mathbb{M}}}^{\underline{\mathbb{M}}}(\Sigma)$ that involves a pair of fundamental $\textrm{E}_{7(7)}$ indices in two different basis: 
\begin{itemize}

\item[$i)$] the index $\mathbb{M}$ in the $\textrm{SL}(8)$ basis decomposing as $\bold{56} \rightarrow \bold{28} + \bold{28}'$, namely, $_{\mathbb{M}} \rightarrow _{[AB]} \oplus ^{[AB]}$ with $A,B=1,..,8$.

\item[$ii)$] the index $\underline{\mathbb{M}}$ in the $\textrm{SU}(8)$ basis decomposing as $\bold{56} \rightarrow \bold{28} + \bold{\overline{28}}$, namely, $_{\underline{\mathbb{M}}} \rightarrow _{[IJ]} \oplus ^{[IJ]}$ with $I,J=1,...,8$. 

\end{itemize}
Since the two types of indices are related by triality when expressed in terms of the common $\textrm{SO}(8)=\textrm{SU}(8) \cap \textrm{SL}(8)$ subgroup, we can swap between the two basis by using invariant tensors $[\gamma^{IJ}]_{AB}$ built as antisymmetric products of gamma matrices in the Majorana-Weyl representation of $\textrm{SO}(8)$. The change of basis is then given by the unitary matrix
\be
\label{Rot_U}
{U_{\underline{\mathbb{M}}}}^{\mathbb{N}} =\frac{1}{2\sqrt{2}}
\left(
\begin{array}{cr}
\left[\gamma_{IJ}\right]^{AB} & i  \left[\gamma_{IJ}\right]_{AB} \\[1mm]
\left[\gamma^{IJ}\right]^{AB} & -i  \left[\gamma^{IJ}\right]_{AB}
\end{array}
\right) \ .
\ee
Using the \textit{mixed} $\textrm{E}_{7(7)}/\textrm{SU}(8)$ coset representative ${\mathcal{V}_{\mathbb{M}}}^{\underline{\mathbb{N}}}(\Sigma)$, one can build the scalar-dependent matrix 
\be
\label{scalar_matrix}
\mathcal{M}_{\mathbb{MN}} = {\mathcal{V}_{\mathbb{M}}}^{\underline{\mathbb{P}}} \,\, {\mathcal{V}_{\mathbb{N}}}^{\underline{\mathbb{Q}}} \,\, \eta_{\underline{\mathbb{PQ}}} 
\hspace{6mm} \textrm{with} \hspace{6mm}
\eta_{\underline{\mathbb{PQ}}} =
\left(
\begin{array}{cr}
0 &\mathbb{I}_{28} \\[1mm]
\mathbb{I}_{28} &0
\end{array}
\right) \ ,
\ee
in terms of which the non-trivial potential induced by the charges ${X_{\mathbb{MN}}}^{\mathbb{P}}$ of the gauged supergravity reads 
\be
\begin{array}{ccl}
\label{V}
V(\mathcal{M}) & = & \dfrac{g^{2}}{672} \left(  {X_{\mathbb{MN}}}^{\mathbb{R}}  {X_{\mathbb{PQ}}}^{\mathbb{S}} \mathcal{M}^{\mathbb{MP}} \mathcal{M}^{\mathbb{NQ}}  \mathcal{M}_{\mathbb{RS}}  +   7 {X_{\mathbb{MN}}}^{\mathbb{Q}} {X_{\mathbb{PQ}}}^{\mathbb{N}} \mathcal{M}^{\mathbb{MP}}  \right) \ .
\end{array}
\ee
This potential is invariant under the linear action of $\textrm{E}_{7(7)}$ transformations and corresponds to the $V$ appearing in (\ref{L_general_gauged}).

\subsection{Fermi mass terms and SUSY transformations}
\label{sec:A1A2_generalities}

The fermionic field content of the supergravity multiplet in maximal supergravity consists of 8 gravitini $\,\psi^{I}_{\mu}\,$ and 56 dilatini $\,\chi_{IJK}=\chi_{[IJK]}\,$. The fermion mass terms in the Lagrangian are given by
\beq
\label{Fermi_Lagrangian}
\begin{array}{cclc}
\mathcal{L}_{\textrm{fermi}} &=&  \frac{\sqrt{2}}{2} \,\, g \, \mathcal{A}_{I J} \,\,\overline{\psi}^{\,\,I}_{\mu} \, \gamma^{\mu \nu} \,\psi^{\,\, J}_{\nu} \,\,\, + \,\,\, \frac{1}{6} \,\, g \, {\mathcal{A}_{I}}^{JKL} \,\,\overline{\psi}^{\,\,I}_{\mu} \, \gamma^{\mu} \, \chi_{JKL} & \\[2mm] 
&+& g \, \mathcal{A}^{IJK,LMN} \,\, \overline{\chi}_{IJK} \,\, \chi_{LMN} \,\,\, + \,\,\, \textrm{h.c.} & ,
\end{array}
\eeq
where $\,\mathcal{A}^{IJK,LMN} \equiv \frac{\sqrt{2}}{144} \, \epsilon^{IJKPQR[LM} \, {\mathcal{A}^{N]}}_{PQR}\,$, and depend on the scalar fields $\,\Sigma_{IJKL}\,$ of the theory. Fermion masses are then totally encoded into the independent tensors $\,{\mathcal{A}^{IJ}=\big( \mathcal{A}_{IJ} \big)^{*}}\,$ and $\,{\mathcal{A}_{I}}^{JKL}=\big( {\mathcal{A}^{I}}_{JKL} \big)^{*}\,$ which transform in the \textbf{36} and the \textbf{420} of $\,\textrm{SU}(8)\,$, after imposing supersymmetry. Similarly to what happened with the embedding tensor in (\ref{QC}), the fermion mass terms are also restricted by the set of quadratic constraints coming from the consistency of the gauging. The above $\mathcal{L}_{\textrm{fermi}}$ is to be identified with the one in (\ref{L_general_gauged}).

%Maximally symmetric solutions are characterised by vanishing vector fields $\,A^{\mathbb{M}}_{\mu}=0\,$ and also by constant scalar configurations $\,\Sigma_{IJKL}=cst\,$ for the $70$ physical scalars of maximal supergravity. The equations of motion can then be written as
%%
%\beq
%\label{scalars_eom}
%\begin{array}{ccc}
%\mathcal{C}_{IJKL} \,+\, \dfrac{1}{4!} \, \epsilon_{IJKLMNPQ} \, \mathcal{C}^{MNPQ} &=& 0 \ ,
%\end{array}
%\eeq
%%
%with $\,\mathcal{C}_{IJKL}={\mathcal{A}^{M}}_{[IJK}\,\mathcal{A}_{L]M} \, +\, \frac{3}{4} \, {\mathcal{A}^{M}}_{N[IJ} \, {\mathcal{A}^{N}}_{KL]M}\,$. Solutions to (\ref{scalars_eom}) correspond to critical points of the scalar potential in (\ref{V}) induced by the gauging, which can be expressed in terms of the (fermi)$^2$ terms as
%%
%%
%\beq
%\label{V_SU8}
%g^{-2} \,V = -\frac{3}{4} \, |\mathcal{A}_1|^{2} + \frac{1}{24} \, |\mathcal{A}_2|^{2} \ ,
%%
%\hspace{10mm} \textrm{ where }\hspace{10mm}
%%
%\left\lbrace
%\begin{array}{ccc}
%|\mathcal{A}_1|^{2}&=&\mathcal{A}_{IJ} \, \, \mathcal{A}^{IJ} \\
%|\mathcal{A}_2|^{2}&=&{\mathcal{A}_{I}}^{JKL} \, {\mathcal{A}^{I}}_{JKL}
%\end{array}
%\right. \ .
%\eeq

The fermion mass terms can be used to compute mass spectra at maximally symmetric solutions (AdS, Minkowski or dS). When evaluated at a critical point of the potential, the scalar spectrum can be computed from
\beq
\label{Mass-matrix_scalars}
\begin{array}{ccl}
g^{-2} \, {\left(\textrm{mass}^{2}\right)_{IJKL}}^{MNPQ}  & =  &  \delta_{IJKL}^{MNPQ} \, \left( \frac{5}{24} \, \mathcal{A}^{R}{}_{STU} \, \mathcal{A}_{R}{}^{STU} - \frac{1}{2} \, \mathcal{A}_{RS} \, \mathcal{A}^{RS} \right) \\[2mm]
& + & 6 \, \delta_{[IJ}^{[MN} \, \left( \mathcal{A}_{K}{}^{RS |P} \, \mathcal{A}^{Q]}{}_{L]RS} - \frac{1}{4} \, \mathcal{A}_{R}{}^{S |PQ]} \, \mathcal{A}^{R}{}_{S|KL]} \right) \\[2mm] 
&-& \frac{2}{3} \, \mathcal{A}_{[I}{}^{[MNP} \, \mathcal{A}^{Q]}{}_{JKL]} \ ,
\end{array}
\eeq
and, as usual in supergravity theories, stability\footnote{For AdS solutions ($\,V_{0}<0\,$) the Breitenlohner-Freedman (B.F.) bound \cite{Breitenlohner:1982jf} for stability does apply. According to it, a solution is stable if $\,m^2 L^2\geq-\frac{9}{4}\,$, where $m^2$ denotes the lowest eigenvalue of the (mass$^2$) matrix and $L^2=-3/V_{0}$ is the AdS radius.} is defined with respect to the normalised mass matrix $\,m^{2} L^2= \frac{3}{|V_{0}|}\,\left(\textrm{mass}^{2}\right)\,$, where $\,V_{0}\,$ denotes the value of the energy in the solution and $L^2=3/|V_{0}|$. The masses of the vectors are obtained after diagonalising the mass matrix
\beq
\label{Mass-matrix_vectors}
g^{-2} \,{\left(\textrm{mass}^{2}\right)_{\underline{\mathbb{M}}}}^{\underline{\mathbb{N}}} = g^{-2} \, \left(
\begin{array}{cc}
{\left(\textrm{mass}^{2}\right)_{IJ}}^{KL} & \left(\textrm{mass}^{2}\right)_{IJKL}  \\
\left(\textrm{mass}^{2}\right)^{IJKL}  & {\left(\textrm{mass}^{2}\right)^{IJ}}_{KL}
\end{array}
\right) \ ,
\eeq
where
\beq
\begin{array}{cclc}
g^{-2} \, {\left(\textrm{mass}^{2}\right)_{IJ}}^{KL}  & =  & -\frac{1}{6} \, {\mathcal{A}_{[I}}^{NPQ} \, \delta_{J]}^{[K} \, {\mathcal{A}^{L]}}_{NPQ}  + \frac{1}{2} \,  {\mathcal{A}_{[I}}^{PQ[K} \, {\mathcal{A}^{L]}}_{J]PQ} & , \\
g^{-2} \, \left(\textrm{mass}^{2}\right)_{IJKL}  & =  & \frac{1}{36} \, {\mathcal{A}_{[I}}^{PQR} \,\epsilon_{J]PQRMNS[K}  \, {\mathcal{A}_{L]}}^{MNS} & .
\end{array}
\eeq
This matrix has (at least) 28 null eigenvalues associated to the unphysical linear combinations of vectors.

Finally, the counterparts of the last term in the modified SUSY transformation (\ref{SUSY-SYM}) are given by
\beq
\delta_{\epsilon} \psi^{\,\, I}_{\mu} = ... +  \sqrt{2} \, g \, \mathcal{A}^{IJ} \gamma_{\mu} \epsilon_{J}
\hspace{10mm} \textrm{,} \hspace{10mm} 
\delta_{\epsilon} \chi^{IJK}= ... -2 \, g \,  {\mathcal{A}_{L}}^{IJK} \epsilon^{L} \ ,
\eeq
where the dots stand for terms already present in the ungauged case \cite{deWit:2007mt}. The number of supersymmetries preserved in a solution corresponds with the number of Killing spinors $\,\epsilon^{J}\,$ satisfying 
\beq
\label{Killing_equations}
g \, \mathcal{A}_{IJ} \, \epsilon^{J} \, = \, \sqrt{-\frac{1}{6}\,V_{0}} \,\, \epsilon_{I} \ .
\eeq
Therefore, the fermion mass terms can be used to thoroughly explore maximally symmetric solutions of supergravity as well as the issues of stability and supersymmetry breaking.

\subsection{The $T$-tensor}
\label{sec:T-Tensor}

The fermion mass terms in (\ref{Fermi_Lagrangian}) can be obtained from the so-called $T$-tensor. This tensor is related to the embedding tensor $\,{X_{\mathbb{MN}}}^{\mathbb{P}}\,$ in (\ref{gauge_brackets}) via a change of basis
\beq
\label{T-Tensor}
{T_{\underline{\mathbb{M}}\underline{\mathbb{N}}}}^{\,\underline{\mathbb{P}}}\,=\,{{\cV}^{\mathbb{M}}}_{\underline{\mathbb{M}}}\,{{\cV}^{\mathbb{N}}}_{\underline{\mathbb{N}}} \, {X_{\mathbb{MN}}}^{\mathbb{P}} \, {{\cV}_{\mathbb{P}}}^{\underline{\mathbb{P}}} \ ,
\eeq
where $\ {{\cV}^{\mathbb{M}}}_{\underline{\mathbb{M}}}\equiv({{\cV^{-1}})_{\underline{\mathbb{M}}}}^{\mathbb{M}}\,$ is the scalar-dependent inverse \textit{mixed} vielbein. The $T$-tensor can then be understood as ``dressing up'' the embedding tensor with the scalar fields of the theory. 

Let us go one step further and decompose the $T$-tensor under the $\,\textrm{SU}(8)\,$ maximal compact subgroup of $\,\textrm{E}_{7(7)}\,$. Applying the index splitting $\,_{\underline{\mathbb{M}}} \rightarrow _{[IJ]} \oplus ^{[IJ]}\,$ to $\,{T_{\underline{\mathbb{M}}\underline{\mathbb{N}}}}^{\,\underline{\mathbb{P}}}\,$ yields various pieces: $\,{T_{IJKL}}^{MN}\,$, $\,T_{IJKLMN}\,$, ... , $\,{T^{IJKL}}_{MN}\,$. Two of them become specially relevant because, upon index contractions (tracing indices), give rise to the fermion mass terms
\beq
\label{tracing}
\begin{array}{lclc}
\mathcal{A}^{IJ}\,=\,\dfrac{4}{21} \, {T^{IKJL}}_{KL} & \hspace{8mm} \textrm{and} \hspace{8mm} & {\mathcal{A}_{I}}^{JKL}\,=\,2 \, {T_{MI}}^{MJKL} & 
\end{array}
\eeq
discussed in the previous section. This is the route we will follow in order to compute the fermion mass terms, scalar/vector masses, etc. later on in the paper.

\subsection{Dyonic gaugings}
\label{sec:Dyonic_generalities}

The possibility to embed a gauging \textit{dyonically} inside the $\textrm{Sp}(56,\mathbb{R})$ electromagnetic group of maximal supergravity was pointed out in refs~\cite{deWit:2002vt,deWit:2005ub,deWit:2007mt,DallAgata:2011aa} and made more concrete in ref.~\cite{Dall'Agata:2012bb}. After this, various gauged supergravities models with $G \subset \textrm{SL}(8)$ have been explored in the literature. This set-up is compatible with choosing electric charges $({X_{\textrm{elec}})_{\mathbb{N}}}^{\mathbb{P}}={\Theta_{[AB]}}^{\textsf{A}} \, {[t_{\mathsf{A}}]_{\mathbb{N}}}^{\mathbb{P}}$ of the form\footnote{In the SL(8) basis of $\textrm{E}_{7(7)}$, the SL(8) generators correspond to block-diagonal generators.}
\be
\label{X_electric}
{X_{[AB] [CD]}}^{[EF]} = -8 \, \delta_{[A}^{[E} \theta_{B][C} \delta_{D]}^{F]} 
\hspace{5mm} , \hspace{5mm}
X_{[AB] \phantom{[CD]} [EF]}^{\phantom{[AB]} [CD]} =   8 \, \delta_{[A}^{[C} \theta_{B][E} \delta_{F]}^{D]} \ ,
\ee
as well as magnetic charges $({X_{\textrm{mag}})_{\mathbb{N}}}^{\mathbb{P}}=\Theta^{[AB] \,  \textsf{A}} \, {[t_{\mathsf{A}}]_{\mathbb{N}}}^{\mathbb{P}}$ given by 
\be
\label{X_magnetic}
X^{[AB] \phantom{[CD]}[EF]}_{\phantom{[AB]}[CD]} = -8 \, \delta_{[C}^{[A} \xi^{B][E} \delta_{D]}^{F]}
\hspace{5mm} , \hspace{5mm}
{X^{[AB] [CD]}}_{[EF]} =   8 \, \delta_{[E}^{[A} \xi^{B][C} \delta_{F]}^{D]}  \ ,
\ee
where the index $\textsf{A}$ is now restricted to run over the $63$ generators of $\textrm{SL}(8)\subset \textrm{E}_{7(7)}$. The symmetric matrices $\theta$ and $\xi$ specify the gauging as a function of the number of positive, negative and vanishing eigenvalues. The set of quadratic constraints in (\ref{QC}) take the form of $\theta \xi = \frac{1}{8} \, \textrm{Tr}(\theta \xi) \, \mathbb{I}_{8} $. Provided $\theta$ is invertible, the solution reads
\be
\label{c-parameter}
\xi = c \, \theta^{-1}
\ee
and allows for a parameter $c$ interpolating between a purely electric gauging at $c=0$ and a purely magnetic one at $c=\infty$. Most of the time it will be more convenient to move to a phase-like parameterisation
\beq
\label{omega-phase}
\omega= \textrm{Arg}(1 + i \, c) \ ,
\eeq
such that purely electric gaugings ($c=0$) correspond to $\omega=0$, purely magnetic ($c=\infty$) to $\omega=\frac{\pi}{2}$ and \textit{dyonic} gaugings to $\omega \in (0,\frac{\pi}{2})$.

In the present paper we will take a second look to the renowned SO(8) gauged supergravity, \textit{i.e.} $\theta=\xi=\textrm{diag(+1, ... ,+1)}$, but will open the door for $\omega \neq 0$ orientations of the gauging inside the electromagnetic $\textrm{Sp}(56,\mathbb{R})$ group.  This selects \textit{dyonic} combinations of vector fields to span the SO(8) gauge symmetry. As mentioned in the introduction, there are the equivalence relations \mbox{$\,\omega \rightarrow -\omega\,$} and \mbox{$\,\omega \rightarrow \omega + \frac{\pi}{4}\,$} for the choice of the electromagnetic phase, hence reducing its relevant range to $\omega \in [0,\frac{\pi}{8}]$.

\section{$\mathcal{N}=2$ truncation}
\label{sec:N=2_trunc}

The dynamics of maximal supergravity results intractable if considering the entire set of fields in the theory. For that reason, it is customary to restrict the field content to a simpler subset invariant under the action of a certain subgroup of the R-symmetry group. We will consider here an SU(3)-invariant sector of the theory whose precise embedding inside the R-symmetry group is given by
\beq
\begin{array}{ccccccccccc}
\textrm{SU}(8) & \rightarrow & \textrm{SO}(8) &  \rightarrow  & \textrm{SO}(7) &\rightarrow & \textrm{G}_{2} \,\,\,\,\textrm{ or }\,\,\,\,\textrm{SU}(4) & \rightarrow & \textrm{SU}(3)  \\[1mm]
\textbf{8} & \rightarrow & \textbf{8}_{v} &  \rightarrow  & \textbf{1} + \textbf{7} &\rightarrow & \textbf{1} +  \textbf{7} \,\,\,\,\,\textrm{ or }\,\,\,\, \textbf{1} + \textbf{1} + \textbf{6} & \rightarrow & \textbf{1} + \textbf{1} + \textbf{3} + \bar{\textbf{3}} 
\end{array}
\eeq
so we decide to identify the $\textbf{8}$ of SU(8) with the $\textbf{8}_{v}$ of the SO(8) gauge group without loss of generality. As a result, the truncated theory contains four vectors -- out of which only two linear combinations are physical -- transforming under the reduced electromagnetic group $\textrm{Sp}(4)$. The gauging in the truncated theory is simply the $\textrm{U}(1)\times \textrm{U}(1)$ commuting with $\textrm{SU}(3)$ inside $\textrm{SO}(8)$. The decomposition of the eight gravitini in maximal supergravity features two singlets revealing the $\mathcal{N}=2$ supersymmetry preserved by the truncation \cite{Bobev:2010ib}. 

The 70 complex scalars in (\ref{SD_condition}) split into self-dual (SD) and anti-self-dual (ASD) irreducible representations (irreps) of SO(8). Schematically, 
\beq
\label{70=35+35}
\textbf{70} = \textbf{35}_{s} \,\textrm{(SD)} + i\, \textbf{35}_{c} \,\textrm{(ASD)} \ .
\eeq
Fields in the $\textbf{35}_{s}$ are proper scalars whereas those in the $\textbf{35}_{c}$ are pseudo-scalars. In the oxidation of the electric ($\omega=0$) SO(8) gauged supergravity to 11d supergravity, the former are related to deformations of the $S^{7}$ metric whereas the latter descend from the antisymmetric $3$-form in the theory. On the other hand, the corresponding operators in the dual field theory are the traceless bosonic and fermionic bilinears, respectively. Given its relevance in this work, we will describe in detail the truncation of the scalar sector.

\subsection{SU(3)-invariant scalars}
\label{sec:SU3_invariant_fields}

Let us denote the components of  a real vector $\vec{x} \in \textbf{8}_{v}$ by $\vec{x}=(x_{1}, ... ,x_{4} \, , \,x_{\hat{1}}, ... ,x_{\hat{4}} )$ and introduce complex variables
\beq
\label{z-complex}
z_{i}=x_{i} + i \, x_{\hat{i}} 
\hspace{6mm} \textrm{ , } \hspace{6mm} 
\bar{z}_{\bar{i}}=x_{i} - i \, x_{\hat{i}}  
\hspace{10mm} \textrm{ with } \hspace{10mm} 
i=1,...,4 \ .
\eeq
These transform as $\,\textbf{4}\,$ and $\,\bar{\textbf{4}}\,$ of $\,\textrm{SU}(4) \subset \textrm{SO}(8)\,$ and have a further $\,``1+3"\,$ splitting
\beq
\label{1+3_spliting}
z_{i}=(\,z_{1} \,,\, z_{a=2,3,4}\,)
\hspace{6mm} \textrm{ , } \hspace{6mm} 
\bar{z}_{\bar{i}}=(\, \bar{z}_{\bar{1}} \,,\, \bar{z}_{\bar{a}=\bar{2},\bar{3},\bar{4}}\, )
\eeq
under $\,\textrm{SU}(3) \subset \textrm{SU}(4)$ with $a$ and $\bar{a}$ transforming in the $\,\bf{3}\,$ and $\,\bar{\bf{3}}\,$ respectively. The self-duality condition for the scalars in (\ref{SD_condition}) is satisfied by the general $\textrm{SU}(3)$-invariant complex four-form
\beq
\label{Sigma-form}
\Sigma = (\sigma_{+} \,\,\Sigma_{+}  +  c.c) \,\,+ \,\, (\sigma_{-} \,\,\Sigma_{-}  +  c.c) \,\,-\,\, \sigma_{R} \,\, J^{+} \wedge J^{+} \,\,-\,\, i \, \sigma_{I} \,\, J^{-} \wedge J^{-} \,\,\, \ ,
\eeq
where $\sigma_{+},\sigma_{-}\in \mathbb{C}$ and $\sigma_{R},\sigma_{I}\in \mathbb{R}$. The basis of invariant forms in (\ref{Sigma-form}) is built using the SU(3)-invariant tensors $\,\left\lbrace \, \delta_{a\bar{a}}\, , \,\epsilon_{abc}\,,\,\epsilon_{\bar{a}\bar{b}\bar{c}} \, \right\rbrace$. These are the two real two-forms
\beq
\label{real-forms}
J^{\pm}=\frac{i}{2}\, \left( \, \pm \, dz_{1} \wedge d\bar{z}_{\bar{1}} \,+\, \sum_{a=1}^{3}  dz_{a} \wedge d\bar{z}_{\bar{a}} \, \right)
\eeq
and the two complex four-forms
\beq
\label{complex-forms}
\Sigma_{+}=dz_{1} \wedge dz_{a} \wedge dz_{b} \wedge dz_{c}
\hspace{6mm} \textrm{ and } \hspace{6mm}
\Sigma_{-}=d\bar{z}_{\bar{1}} \wedge dz_{a} \wedge dz_{b} \wedge dz_{c} \ ,
\eeq
together with the conjugates $\Sigma_{+}^{*}$ and $\Sigma_{-}^{*}$.  Inserting (\ref{real-forms}) and (\ref{complex-forms}) into (\ref{Sigma-form}) and plugging the complex variables in (\ref{z-complex}), one can read off the components of $\Sigma$ using the original coordinates $(x_{i} , x_{\hat{i}} )$.

The scalar fields $\sigma_{R},\sigma_{I}\in \mathbb{R}$ and $\sigma_{+},\sigma_{-}\in \mathbb{C}$ in the SU(3)-truncation of maximal supergravity describe the coset space $\,\mathcal{M}_{\textrm{scalar}}=\frac{\textrm{SL}(2)}{\textrm{SO}(2)} \times \frac{\textrm{SU}(2,1)}{\textrm{SU}(2) \times \textrm{U}(1)}$. It contains two factors which are respectively the special K\"ahler (SK) and quaternionic K\"ahler (QK) manifolds in the $\mathcal{N}=2$ truncated theory. The two supersymmetries are associated to the $\psi_{\mu}^{1}$ and $\psi_{\mu}^{\hat{1}}$ gravitini which are singlets under the SU(3) action. It becomes very convenient to introduce a set of new variables 
\beq
\label{Fields_of_V}
\begin{array}{cclclc}
\varpi &=& \phantom{-} \sigma_{R} + i \, \sigma_{I} & = & \lambda \,  e^{i\alpha} & , \\[1mm]
\varpi_{1} &=& \phantom{-} \text{Re}(\sigma_{+}) + i \, \text{Im}(\sigma_{-}) & = & \lambda' \,  \left( \, e^{i \phi} \cos\theta \cos\psi - e^{-i \phi} \sin\theta \sin\psi \, \right) & , \\[1mm]
\varpi_{2} &=& -\text{Im}(\sigma_{+}) + i \, \text{Re}(\sigma_{-}) & = & \lambda' \,  \left( \, e^{i \phi} \cos\theta \sin\psi + e^{-i \phi} \sin\theta \cos\psi \, \right) & ,
\end{array}
\eeq
which amounts to an alternative expansion
\beq
\label{new_expansion}
\begin{array}{ccl}
\Sigma & = & \textrm{Re}(\varpi )\, J^{+} \wedge J^{+} + i \, \textrm{Im}(\varpi )\,  J^{-} \wedge J^{-} \\[1mm]
            & + & \textrm{Re}(\varpi_{1} ) \, \textrm{Re}(\Sigma_{+}) + i \, \textrm{Im}(\varpi_{1} ) \, \textrm{Re}(\Sigma_{-}) \,\,+\,\, \textrm{Re}(\varpi_{2} ) \, \textrm{Im}(\Sigma_{+}) + i \, \textrm{Im}(\varpi_{2} ) \, \textrm{Im}(\Sigma_{-}) \ .
\end{array}
\eeq
Using this expansion, $\textrm{Re}(\varpi )$, $\textrm{Re}(\varpi_{1} )$ and  $\textrm{Re}(\varpi_{2} )$ correspond to scalars in the $\textbf{35}_{s}$ whereas $\textrm{Im}(\varpi )$, $\textrm{Im}(\varpi_{1} )$ and  $\textrm{Im}(\varpi_{2} )$ correspond to pseudo-scalars in the $\textbf{35}_{c}$, in agreement with the splitting (\ref{70=35+35}). 

The complex scalar $\varpi=\lambda \,  e^{i\alpha}$ parameterises the SK manifold whereas $(\varpi_{1},\varpi_{2})$ parameterises the QK manifold in terms of the modulus $\lambda'$ and the three SU(2) phases $\phi$ and $(\theta , \psi)$. The main advantage of this parameterisation is that the $\textrm{U}(1) \times \textrm{U}(1)$ gauge symmetry in the truncated theory can be used to gauge-fixing $\,\theta=\psi=0\,$ \cite{Warner:1983vz,Ahn:2000mf,Bobev:2010ib}. This translates into $\varpi_{1}=\lambda' e^{i \phi}$ and $\varpi_{2}=0$, so that we are left with a theory containing four real scalars $(\lambda , \alpha)$ and $(\lambda' , \phi)$. Furthermore, this gauge choice implies that there are no four-forms with an odd number of hatted (unhatted) indices in the expansion (\ref{new_expansion}), \textit{e.g.} $\,\Sigma_{\hat{1}234}\,$ , $\,\Sigma_{1\hat{2}\hat{3}\hat{4}}\,$, etc., since they only appear through $\textrm{Im}(\Sigma_{+})$ and $\textrm{Im}(\Sigma_{-})$. In the absence of these ``odd" forms, the truncated theory admits an intermediate $\mathcal{N}=4$ formulation \`a la Sch\"on$\&$Weidner \cite{Schon:2006kz} that makes a connection to generalised type II flux compactifications feasible \cite{Dibitetto:2012ia}. We would like to look into this in the future.

\subsection{The scalar Lagrangian}
\label{sec:SU3_Lagrangian}

The Lagrangian for the scalar sector of maximal supergravity is given by
\be
\label{Lscalar}
\mathcal{L}_{\textrm{scalar}} = \frac{1}{96} \textrm{Tr}(D_{\mu} \mathcal{M} \,\, D^{\mu} \mathcal{M}^{-1}) - V(\mathcal{M}) \ ,
\ee
where $\,\mathcal{M}\equiv \mathcal{M}_{\mathbb{MN}}\,$ is the scalar-dependent matrix in (\ref{scalar_matrix}) built from the \textit{mixed} vielbein $\,{\mathcal{V}_{\mathbb{M}}}^{\underline{\mathbb{P}}}(\lambda,\alpha,\lambda',\phi,\theta,\psi)\,$. At this point we are not performing any gauge-fixing yet, so we deal with a six real fields problem. The construction of the vielbein depends on the specific choice of basis for the $\textrm{E}_{7(7)}$ generators and other related issues. In order to keep this section alive, we have put all the details aside in the appendix~\ref{App:E7_parameterisaton}. 

The covariant derivative induced by the gauging is totally encoded inside the $\omega$-dependent embedding tensor in (\ref{X_electric})-(\ref{X_magnetic}) and reads
\beq
D_{\mu} \mathcal{M}_{\mathbb{MN}} = \partial_{\mu} \mathcal{M}_{\mathbb{MN}} - 2 \, g \, A_{\mu}^{\mathbb{P}} \, \, {X_{\mathbb{P}(\mathbb{M}}}^{\mathbb{Q}} \, \mathcal{M}_{\mathbb{N})\mathbb{Q}} \ .
\eeq
In this work we will consider vanishing vector fields $\,A^{M}_{\mu}=0\,$ compatible with maximally symmetric solutions of the theory and also with BPS domain-wall configurations interpolating between two of such solutions. As a consequence $\,D_{\mu} \rightarrow \partial_{\mu}\,$ and the scalar Lagrangian takes the form
\beq
\label{L_scalar_SU3}
\begin{array}{ccl}
\mathcal{L}_{\textrm{scalar}} &=& -3\, (\partial_{\mu} \lambda)^2 - \frac{3}{4}\,  \sinh^2(2 \lambda)\,  (\partial_{\mu} \alpha)^2 - 4\, (\partial_{\mu} \lambda')^2 - \sinh^2(2 \lambda')\,  (\partial_{\mu} \phi)^2 \\ 
&-& T(\lambda',\phi,\theta,\psi) - V(\lambda,\alpha,\lambda',\phi,\theta,\psi) \ ,
\end{array}
\eeq
where
\beq
\label{T_kinetic}
\begin{array}{ccl}
T(\lambda',\phi,\theta,\psi) &=& \left[ \sinh ^4(2 \lambda') \cos ^2(2 \phi )+\sinh ^2(2 \lambda') \right]\, (\partial_{\mu} \theta)^2  +  \frac{1}{4} \sinh ^2(4 \lambda') \, (\partial_{\mu} \psi)^2 \\[1mm]
&+& \frac{1}{2} \sinh ^2(4 \lambda') \cos (2 \phi ) \, (\partial_{\mu} \theta) (\partial_{\mu} \psi) \ ,
\end{array}
\eeq
accounts for the kinetic energy associated to the fields $(\theta,\psi)$ which, as discussed before, can be gauged away.

The computation of the scalar potential $V(\lambda,\alpha,\lambda',\phi,\theta,\psi)$ for a \textit{dyonic} gauging turns out to be rather cumbersome mostly due to the cubic term $XX\mathcal{M}\mathcal{M}\mathcal{M}$ in (\ref{V}). To carry it out, it is helpful to use the parameter $c$ in (\ref{c-parameter}) instead of its compact version $\omega$ in (\ref{omega-phase}). We set the normalisation with an overall factor $1/(1+c^2)$. After a straightforward but tedious computation, the $c$-dependent scalar potential in (\ref{L_scalar_SU3}) reads
\beq
\label{VSU3}
\begin{array}{ccl}
&V&\hspace{-3mm}(\lambda,\alpha,\lambda',\phi) \,\,= \\[3mm]
&=& \dfrac{g^2}{128 \left(1+c^2\right)} \Big[ 4\,  \Big( (c^2+1) \cosh (6 \lambda ) \sinh ^2(2 \lambda ') \,  (19 \cosh (4 \lambda ')+21) \\[3mm]
&-&4 \sinh (2 \lambda ) \Big(2 \sinh ^2(2 \lambda ) \cos (4 \phi ) \sinh ^4(2 \lambda ') \Big( (c^2-1) \cos (3 \alpha )-2 c \sin (3 \alpha ) \Big) \\[2mm]
&+& \sinh ^2(2 \lambda ') \Big( 3 (c^2-1) \cos (\alpha ) \Big( \cosh (4 \lambda ) (3 \cosh (4 \lambda ')+2 \cos (2
   \phi )+3) \\[2mm]
&+& \cosh (4 \lambda ')-6 \cos (2 \phi )-7 \Big)+\sinh ^2(2 \lambda ) \, (\cosh (4 \lambda ')+3)
   \Big( (c^2-1) \cos (3 \alpha )-2 c \sin (3 \alpha ) \Big) \\[2mm]
&+& 6 \, c \, \sin (\alpha ) \Big(\cosh (4 \lambda ')-2 \, (\cosh (4 \lambda )-3) \cos (2\phi )-7 \Big) \Big) \\[2mm]
&+& 3 \sinh ^2(4 \lambda ') \Big(3 c \sin (\alpha ) \cosh (4 \lambda )-(c^2+1) \cos (2 \alpha ) \sinh (4 \lambda )
   \cos (2 \phi )\Big) \Big) \Big) \\[2mm]
&+& 32\, (c^2+1) \cosh ^3(2 \lambda ) \cos (4 \phi ) \sinh ^4(2 \lambda ') \\[2mm]
&+& 3\, (c^2+1) \cosh (2  \lambda ) \, \Big(3 \, (\cosh (8 \lambda ')-45)-124 \cosh (4 \lambda ') \Big)\\[2mm]
&-&192 \, \sinh (2 \lambda ) \cosh ^2(2 \lambda ) \cos (2 \phi ) \sinh ^2(2 \lambda ') \cosh (4 \lambda ') \Big( (c^2-1) \cos (\alpha )-2 c \sin (\alpha )\Big) \Big] \ .
\end{array}
\eeq
The \textit{dyonic} potential does not depend on the fields $(\theta,\psi)$ which can be gauged-away at any value of $\,c\,$, in analogy to the  purely electric gauging $c=0$ studied in ref.~\cite{Ahn:2009as}. The reason is that the $c$ parameter encodes an $\textrm{Sp}(56)$ rotation that does not modify the embedding $\,\textrm{SU}(3) \subset \textrm{SO}(8)\,$.  On the other hand, the above scalar potential is invariant under the transformations $\phi \rightarrow \phi + \pi$ and $\phi \rightarrow -\phi$. The latter will be connected later to the existence of two different superpotentials in the $\mathcal{N}=2$ truncated theory.

\subsection{The fermi mass terms}
\label{sec:SU3_fermi_square_terms}

In this section we compute the fermi mass terms $\mathcal{A}^{IJ}$ and ${\mathcal{A}_{I}}^{JKL}$ in (\ref{Fermi_Lagrangian}) as a function of the scalars (after gauge-fixing $\theta=\psi=0$) and the electromagnetic phase $\omega$ in (\ref{omega-phase}). To do so, we follow the prescription described in section~\ref{sec:T-Tensor} : first we build the $T$-tensor in (\ref{T-Tensor}) using the explicit form of the \textit{mixed} vielbein $\,{{\cV}_{\mathbb{P}}}^{\underline{\mathbb{P}}}(\lambda, \alpha, \lambda',\phi)\,$ and then extract $\mathcal{A}^{IJ}$ and ${\mathcal{A}_{I}}^{JKL}$ by taking the traces in (\ref{tracing}).

\subsubsection{Gravitino-gravitino terms}
\label{sec:A1_terms}

The computation of the gravitino-gravitino couplings $\mathcal{A}^{IJ}(\lambda,\alpha,\lambda',\phi)$ reveals an splitting of the the $\omega$-dependence of the form
\beq
\label{A1_splitting}
\mathcal{A}^{IJ}=e^{i \omega} \, \mathcal{A}^{IJ}_{+} + e^{-i \omega} \, \mathcal{A}^{IJ}_{-} \ .
\eeq
Recalling the index decomposition $\,I \rightarrow 1 \oplus a \oplus \hat{1} \oplus \hat{a}\,$, the mass terms for the two gravitini which are singlets under SU(3) and therefore survive the truncation to the $\mathcal{N}=2$ theory read
\beq
\begin{array}{cclc}
\label{A11_mass_term}
\mathcal{A}^{11}_{+} &=&\tfrac{3}{2} e^{ i (2 \alpha +2 \phi )} \cosh (\lambda ) \sinh ^2(\lambda )  \sinh ^2(2 \lambda' )+\cosh ^3(\lambda ) \, f_{1}(\lambda',\phi)  & , \\[2mm]
\mathcal{A}^{11}_{-} &=&\tfrac{3}{2} e^{i (\alpha +2 \phi )} \sinh (\lambda ) \cosh^2(\lambda ) \sinh ^2(2 \lambda' ) + e^{3 i \alpha }  \,  \sinh ^3(\lambda ) \, f_{1}(\lambda',\phi) & , 
\end{array}
\eeq
together with
\beq
\label{A1hat1hat_mass_term}
\mathcal{A}^{\hat{1}\hat{1}}_{\pm}(\lambda,\alpha,\lambda',\phi)=\mathcal{A}^{11}_{\pm}(\lambda,\alpha,\lambda',-\phi) \ . 
\eeq
The remaining six non-singlet gravitini which are projected out in the truncated theory acquire a mass term
\beq
\label{Aaa_mass_term}
\begin{array}{ccr}
\mathcal{A}^{aa}_{+} &=&\frac{1}{8} \cosh (\lambda ) \Big[ 4 \, e^{-2 i \alpha } \sinh ^2(\lambda ) \sinh ^2(2 \lambda' ) \cos (2 \phi )+\cosh (2 \lambda ) \, g_{1}(\lambda') -\cosh (4 \lambda' )+5 \Big] \ , \\[3mm]
\mathcal{A}^{aa}_{-}&=& \tfrac{e^{-i \alpha }}{8} \sinh (\lambda ) \Big[ 4 \, e^{2 i \alpha } \cosh ^2(\lambda ) \sinh ^2(2 \lambda' ) \cos (2 \phi )+\cosh (2 \lambda )\,  g_{1}(\lambda') +\cosh (4 \lambda' )-5 \Big] \ , \\[3mm]
\end{array}
\eeq
together with
\beq
\label{Aahatahat_mass_term}
\mathcal{A}^{\hat{a}\hat{a}}_{\pm}(\lambda,\alpha,\lambda',\phi)=\mathcal{A}^{aa}_{\pm}(\lambda,\alpha,\lambda',\phi) \ .
\eeq
In order to shorten the above expressions, as well as some forthcoming ones, we have introduced the functions
\beq
\begin{array}{rclcrclc}
f_{1}(\lambda',\phi)  &=&  \cosh ^4(\lambda' )+e^{4 i \phi } \sinh ^4(\lambda' ) & , &
g_{1}(\lambda')  &=&  3 \cosh (4 \lambda' )+1 & .
\end{array}
\eeq
As a check of consistency, the expressions in refs~\cite{Ahn:2000mf,Ahn:2009as} for the pure electric SO(8) gauging are exactly recovered\footnote{By redefining the fields as $p=\cosh(\lambda)$ , $q=\sinh(\lambda)$ , $r=\cosh(\lambda')$ and $t=\sinh(\lambda')$, the mass term $\mathcal{A}^{11}$ in (\ref{A1_splitting}) is written as
\beq
\mathcal{A}^{11}=e^{i\, \omega} \, \Big(   p^3 \left(r^4+t^4 e^{4 i \phi }\right)+6 p \, q^2\, r^2\, t^2 e^{2 i (\alpha +\phi )}      \Big)  + e^{-i\, \omega}\, \Big( e^{3 i \alpha } q^3 \left(r^4+t^4 e^{4 i \phi }\right) +    6 q \,  p^2 \,  r^2 \, t^2 e^{i (\alpha +2 \phi )}     \Big)  \ , \nonumber
\eeq
which exactly reproduces the expression (2.29) in ref.~\cite{Ahn:2009as} when $\omega=0$. The rest of the fermion mass terms also match precisely if setting $\omega=0$. 
} by setting $\omega=0$. 

\subsubsection{Gravitino-dilatino terms}
\label{sec:A2_terms}

An explicit computation of the $\,{\mathcal{A}_{I}}^{JKL}\,$ tensor shows once more a simple $\omega$-dependence of the form 
\beq
\label{A2_splitting}
{\mathcal{A}_{I}}^{JKL} \,\,=\,\,  {\mathcal{A}_{+I}}^{JKL} \, e^{ i \omega} \,\,+\,\, {\mathcal{A}_{-I}}^{JKL} \, e^{ -i \omega} \ ,
\eeq
as for the gravitino-gravitino mass terms (\ref{A1_splitting}).  Moreover, because of the gauge choice $\theta = \psi = 0$, all the components involving an odd number of hatted (unhatted) indices vanish: ${\mathcal{A}_{\hat{1}}}^{abc}=0$ , ${\mathcal{A}_{a}}^{a b \hat{c}}=0$ , etc. In order to present the different terms, it is again convenient to organise the fermions according to the index decomposition $\,I \rightarrow 1 \oplus a \oplus \hat{1} \oplus \hat{a}\,$. Below we just list those fermions for which a fermi-fermi coupling is generated
\beq
\label{fermi_couplings}
\begin{array}{cccc}
\textrm{gravitini} & \hspace{5mm} & \textrm{dilatini} \\[2mm]
\psi^{1}_{\mu}  &  &   \chi_{abc}  \,\,\,\, , \,\,\,\,  \chi_{\hat{1} a \hat{a}}  \,\,\,\, , \,\,\,\, \chi_{a\hat{b}\hat{c}}   & , \\[2mm]
\psi^{a}_{\mu}  & &  \chi_{1\hat{b}\hat{c}}  \,\,\,\, , \,\,\,\,  \chi_{\hat{a} 1 \hat{1}}  \,\,\,\, , \,\,\,\, \chi_{\hat{a} b \hat{b}}   \,\,\,\, , \,\,\,\, \chi_{\hat{1} b \hat{c}}  \,\,\,\, , \,\,\,\, \chi_{1 b c}  & , \\[2mm]
\psi^{\hat{1}}_{\mu} & & \chi_{\hat{a} \hat{b} \hat{c}}  \,\,\,\, , \,\,\,\, \chi_{1a\hat{a}}   \,\,\,\, , \,\,\,\, \chi_{a b\hat{c}}   & , \\[2mm]   
\psi^{\hat{a}}_{\mu}  & & \chi_{\hat{1} b c}  \,\,\,\, , \,\,\,\,  \chi_{a 1 \hat{1}}  \,\,\,\, , \,\,\,\, \chi_{a b \hat{b}}   \,\,\,\, , \,\,\,\, \chi_{1 b \hat{c}}  \,\,\,\, , \,\,\,\, \chi_{\hat{1} \hat{b} \hat{c}}   & ,\\[2mm]
\end{array}
\eeq
where the first gravitino only couples to the first row of dilatini, the second to the second row and so on.  As an example, there is a mass term $\,\mathcal{L}_{\textrm{fermi}} \supset \frac{1}{6} \,  g \, {\mathcal{A}_{a}}^{\hat{a} 1\hat{1}} \, \bar{\psi}^{a}_{\mu} \,\,  \chi_{\hat{a} 1 \hat{1}}\,$ given by
\beq
\label{A2_example}
\begin{array}{cclcc}
{\mathcal{A}_{+a}}^{\hat{a} 1\hat{1}} & = &\,\,\,\,\,\,\,\,\,  -\frac{1}{4}  \cosh (\lambda ) \Big[  2 \cosh^2(\lambda ) \sinh ^2(2 \lambda' ) \cos(2 \phi) + e^{2 i \alpha} \sinh ^2(\lambda ) g_{1}(\lambda') \Big]  & ,\\[3mm]
{\mathcal{A}_{-a}}^{\hat{a} 1\hat{1}} & = & -\frac{e^{3 i \alpha }}{4}  \sinh (\lambda ) \Big[  2 \, \sinh ^2(\lambda ) \sinh ^2(2 \lambda' ) \cos (2 \phi)  +  e^{-2 i \alpha }\cosh ^2(\lambda )  g_{1}(\lambda') \Big] & .
\end{array}
\eeq
The complete set of non-vanishing gravitino-dilatino couplings is listed in appendix~\ref{App:fermi_masses}.  Knowing all the fermion mass terms in (\ref{Fermi_Lagrangian}) will allow us to compute the full $\mathcal{N}=8$ mass spectra at any critical point of the scalar potential (\ref{VSU3}) by using the mass formulae (\ref{Mass-matrix_scalars}) and (\ref{Mass-matrix_vectors}).

\subsection{$\mathcal{N}=2$ superpotentials}
\label{sec:superpotentials}

Due to the $\mathcal{N}=2$ supersymmetry preserved by the SU(3)-truncation, there exist two superpotentials, we will denote by $W_{1}$ and $W_{\hat{1}}$, from which the scalar potential in (\ref{VSU3}) can be derived. The $W_{1}$ and $W_{\hat{1}}$ superpotentials are identified with the $\mathcal{A}^{11}$ and $\mathcal{A}^{\hat{1}\hat{1}}$ mass terms of the two SU(3)-singlet gravitini in (\ref{A1_splitting}) \cite{Ahn:2009as}. As a consequence, they depend on the fields $(\lambda, \alpha, \lambda' , \phi)$ as well as on the electromagnetic parameter $\omega$, namely, 
\beq
\label{W1&What1}
W_{1}=e^{i \omega} \, \mathcal{A}^{11}_{+} + e^{-i \omega} \, \mathcal{A}^{11}_{-}
\hspace{10mm} \textrm{ or } \hspace{10mm}
W_{\hat{1}}=e^{i \omega} \, \mathcal{A}^{\hat{1}\hat{1}}_{+} + e^{-i \omega} \, \mathcal{A}^{\hat{1}\hat{1}}_{-} \ .
\eeq
Looking at the form of (\ref{A11_mass_term}), it is easy to see that both superpotentials remain invariant under the shift $\phi \rightarrow \phi+ \pi$ and, by virtue of (\ref{A1hat1hat_mass_term}), are exchanged by the reflection $\phi \rightarrow -\phi$. Using any of the two complex superpotentials above,
\beq
\label{W_choice}
W=W_{1}(\lambda, \alpha, \lambda' , \phi)
\hspace{10mm} \textrm{ or } \hspace{10mm}
W=W_{\hat{1}}(\lambda, \alpha, \lambda' , \phi) \ , 
\eeq
the scalar potential can be derived as\footnote{The different coefficients with respect to refs~\cite{Ahn:2000mf,Ahn:2009as} stem from a different normalisation: ${\lambda_{\textrm{here}}=\lambda_{\textrm{there}}/2\sqrt{2}}$ and $\lambda'_{\textrm{here}}=\lambda'_{\textrm{there}}/2\sqrt{2}$ . }
\beq
\label{V_from_W}
\begin{array}{ccl}
V(\lambda,\alpha,\lambda',\phi) &=& g^2 \, \left[  \, \dfrac{2}{3}   \, \left| \partial_{\lambda} W\right|^2 \, + \, \dfrac{1}{2}  \,\left| \partial_{\lambda'} W \right|^2  - 6\,  \left| W \right|^2  \right] \\[4mm]
 & = &  g^2 \, \left[  \, \dfrac{2}{3}   \,  \left(\partial_{\lambda} |W|\right) ^2 \, + \, \dfrac{8}{3  \sinh^2(2 \lambda)}   \,  \left(\partial_{\alpha} |W|\right) ^2 \right.  \\[4mm] 
 &+&  \left. \phantom{g^2}\,\,\,\,\,\dfrac{1}{2}  \, \left(\partial_{\lambda'} |W|\right) ^2   +  \dfrac{2}{\sinh^2(2 \lambda')}    \left(\partial_{\phi} |W|\right) ^2 - 6\,  | W |^2  \right] \ .
\end{array}
\eeq
In going from the first line to the second in (\ref{V_from_W}) we write $W=|W|e^{i \textrm{Arg}(W)}$ and use the relations
\beq
\label{constraintsW}
\begin{array}{lclc}
|W| \, \partial_{\lambda} \textrm{Arg}(W) & = & - \dfrac{2}{\sinh(2 \lambda)} \,  \partial_{\alpha} |W| & , \\[3mm]
|W| \, \partial_{\lambda'} \textrm{Arg}(W) & = & \mp \dfrac{2}{\sinh(2 \lambda')} \,  \partial_{\phi} |W|  & . \\
\end{array}
\eeq
It is straightforward to check that the $\omega$-dependent superpotentials $W_{1}$ and $W_{\hat{1}}$ in (\ref{W_choice}) satisfy the conditions (\ref{constraintsW}) for the upper and lower sign choice respectively, and that the scalar potential computed from (\ref{V_from_W}) by plugging (\ref{W1&What1}) exactly matches the one in (\ref{VSU3}) computed from (\ref{V}). The real and $\omega$-dependent function $|W(\lambda, \alpha, \lambda' , \phi)|$ will become the relevant one when looking at BPS domain-wall configurations in the next section.

Let us now introduce new complex variables
\beq
\label{field-redef}
z = \tanh(\lambda) \, e^{i \alpha}
\hspace{10mm} \textrm{ and } \hspace{10mm}
\zeta_{12} = \tanh(\lambda') \, e^{i \phi} \ .
\eeq
Using the form of the gravitino-gravitino mass terms in (\ref{A11_mass_term}), and after some algebra manipulations, the $\,W_{1}\,$ superpotential in (\ref{W1&What1}) takes the form 
\beq
\label{W1z-zeta}
W_{1}(z,\zeta_{12})=\frac{ \left(e^{i \omega }+e^{-i \omega } z^3\right) \, \left( 1 + \zeta_{12}^4 \right)  + 6 \,  z \, \left( e^{-i \omega } + e^{i \omega } z\right) \, \zeta_{12}^2}{ (1-|z|^2)^{\frac{3}{2}} \,\, (1-|\zeta_{12}|^2)^2} \ .
\eeq
The above superpotential represents the generalisation to arbitrary values of $\omega$ of the one derived in ref.~\cite{Bobev:2010ib}, which now we know corresponds to $\,\omega=0$. A \textit{conjectured} $\omega$-dependent superpotential was first presented in ref.~\cite{Borghese:2012zs}. Even though the generalisation hinged on symmetry arguments\footnote{It was based on the invariant classifiers computed in ref.~\cite{Dall'Agata:2012bb}.} involving the periodicity of $\omega$, a full-fledged supergravity derivation of the $\omega$-dependent superpotential was missing. Here we have provided such a derivation using the framework of the embedding tensor, finding that the conjectured superpotential in ref.~\cite{Borghese:2012zs} was correct up to an overall phase that could not be determined by symmetry arguments therein.

\section{Scalar dynamics and BPS domain-walls}
\label{sec:scalar_dynamics_BPS}

The dynamics of the SU(3)-invariant scalar sector is encoded in the action
\beq
\label{S_scalar}
S_{\textrm{scalar}} = \int d^{4}x \sqrt{-g} \left( \, \frac{1}{2} \, R  - \frac{1}{2} \, K_{ij} (\partial_{\mu} \Sigma^{i})  (\partial^{\mu} \Sigma^{j}) - V(\Sigma^{i}) \, \right) \ ,
\eeq
where we have collectively denoted $\Sigma^{i}=(\lambda, \alpha, \lambda',\phi)$. The field-space metric $K_{ij}$ can be read off from (\ref{L_scalar_SU3}) finding
\beq
\label{K-metric}
K_{ij} = 
\left(
\begin{array}{cccc}
 6 & 0 & 0 & 0 \\
 0 & \frac{3}{2} \sinh ^2(2 \lambda ) & 0 & 0 \\
 0 & 0 & 8 & 0 \\
 0 & 0 & 0 & 2 \sinh ^2(2 \lambda') \\
\end{array}
\right) \ ,
\eeq
and the scalar potential $V(\Sigma^{i})$ was given in (\ref{VSU3}) (alternatively (\ref{V_from_W})). We will make a domain-wall Ansatz for the space-time metric
\beq
\label{g_DW}
ds^2 = e^{2 A(z)} \, \eta_{\alpha \beta}   \, dx^{\alpha} dx^{\beta} \,\,+\,\, dz^2
\hspace{5mm}  \textrm{ with } \hspace{5mm}
\eta_{\alpha \beta}=\textrm{diag}(-1,+1,+1) \ ,
\eeq
where $z\in (-\infty,\infty)$ is the coordinate transverse to the domain-wall and $A(z)$ is the scale factor. 

The non-vanishing components of the Einstein equations $G_{\mu\nu}=T_{\mu \nu}$ obtained from the action (\ref{S_scalar}) read
\beq
\begin{array}{rcrl}
\label{Einstein_EOM}
3 \, (\partial_{z}A)^2 + 2 \, \partial^2_{z}A &=& - \frac{1}{2} \, K_{ij} (\partial_{z} \Sigma^{i}) (\partial_{z} \Sigma^{j}) - V(\Sigma) & , \\
3 \, (\partial_{z}A)^2                &=& \frac{1}{2} \, K_{ij} (\partial_{z} \Sigma^{i}) (\partial_{z} \Sigma^{j}) - V(\Sigma) & .
\end{array}
\eeq
These two equations can be combined to obtain the simple monotonicity relation
\beq
\label{A'_monotonic}
\partial^2_{z} A = - \frac{1}{2} \, K_{ij} (\partial_{z} \Sigma^{i}) (\partial_{z} \Sigma^{j})  \le 0 \ ,
\eeq
so that $\partial_{z}A$ will decrease along the domain-wall solution. The Euler-Lagrange equations for the scalars 
\beq
\label{Euler-Lagrange_general}
\square \Sigma^{i} + \left[  \partial_{\rho} g^{\rho \mu}   + \Gamma^{\nu}_{\nu \rho} \, g^{\rho \mu}    \right] (\partial_{\mu} \Sigma^{i}) + \Gamma^{i}_{jk} (\partial_{\rho} \Sigma^{j}) (\partial^{\rho} \Sigma^{k}) - K^{ij} (\partial_{j}V) = 0 \ ,
\eeq
with $\Gamma^{\mu}_{\nu \rho}$ and $\Gamma^{i}_{jk}$ denoting Christoffel symbols in space-time and field-space, give rise to the following equations of motion :
%
%\begin{eqnarray}
%\label{Euler-Lagrange_fields}
%\square \lambda + 3 \, (\partial_{z} A) \, (\partial_{z} \lambda) - \frac{1}{4} \, \sinh(4\lambda)  \, (\partial_{z} \alpha)^2 - \frac{1}{6} \, \partial_{\lambda} V &=& 0 \ , \nonumber \\
%\sinh^2(2 \lambda) \, \square \alpha + 3 \, \sinh^2(2 \lambda) \, (\partial_{z}A) \, (\partial_{z} \alpha) + 2 \,  \sinh(4\lambda)  \, (\partial_{z} \alpha) \, (\partial_{z} \lambda) - \frac{2}{3} \, \partial_{\alpha} V &=& 0 \ , \nonumber \\
%%
%\\[-4mm]
%\square \lambda' + 3 \, (\partial_{z} A) \, (\partial_{z} \lambda') - \frac{1}{4} \, \sinh(4\lambda')  \, (\partial_{z} \phi)^2 - \frac{1}{8} \, \partial_{\lambda'} V &=& 0 \ , \nonumber \\
%\sinh^2(2 \lambda') \, \square \phi + 3 \, \sinh^2(2 \lambda') \, (\partial_{z} A) \, (\partial_{z} \phi) + 2 \,  \sinh(4\lambda')  \, (\partial_{z} \phi) \, (\partial_{z} \lambda') - \frac{1}{2} \, \partial_{\phi} V &=& 0 \ , \nonumber 
%\end{eqnarray}
%
%
\begin{eqnarray}
\label{Euler-Lagrange_fields}
0  &=& \square \lambda + 3 \, (\partial_{z} A) \, (\partial_{z} \lambda) - \frac{1}{4} \, \sinh(4\lambda)  \, (\partial_{z} \alpha)^2 - \frac{1}{6} \, \partial_{\lambda} V   \nonumber \\
0  &=& \sinh^2(2 \lambda) \, \square \alpha + 3 \, \sinh^2(2 \lambda) \, (\partial_{z}A) \, (\partial_{z} \alpha) + 2 \,  \sinh(4\lambda)  \, (\partial_{z} \alpha) \, (\partial_{z} \lambda) - \frac{2}{3} \, \partial_{\alpha} V  \nonumber \\
\\[-4mm]
0  &=& \square \lambda' + 3 \, (\partial_{z} A) \, (\partial_{z} \lambda') - \frac{1}{4} \, \sinh(4\lambda')  \, (\partial_{z} \phi)^2 - \frac{1}{8} \, \partial_{\lambda'} V  \nonumber \\
0  &=& \sinh^2(2 \lambda') \, \square \phi + 3 \, \sinh^2(2 \lambda') \, (\partial_{z} A) \, (\partial_{z} \phi) + 2 \,  \sinh(4\lambda')  \, (\partial_{z} \phi) \, (\partial_{z} \lambda') - \frac{1}{2} \, \partial_{\phi} V  \nonumber 
\end{eqnarray}

We will obtain AdS solutions to the above system of equations as well as BPS domain-wall configurations which additionally satisfy first-order flow equations.

\subsection{AdS solutions}

Maximally symmetric solutions are characterised by scalar fields getting a constant vacuum expectation value (VEV), \textit{i.e.} $\partial_{\mu} \Sigma^{i}=0$. The equations in (\ref{Euler-Lagrange_fields}) boil down to extremisation conditions
\beq
\label{critcal_point}
\partial_{\lambda} V  = \partial_{\alpha} V = \partial_{\lambda'} V = \partial_{\phi} V = 0 \ ,
\eeq
and the Einstein equations reduce to $G_{\mu \nu} + V_{0} \, g_{\mu \nu} = 0 $. The cosmological constant $V_{0}$ is just the scalar potential evaluated at the critical point. The space-time metric then becomes that of Anti-de Sitter (AdS), Minkowski (Mkw) or de Sitter (dS) space for $V_{0}<0$, $V_{0}=0$ and $V_{0}>0$, respectively. In the case of AdS, which is the relevant in this paper, the solution to the scale factor in (\ref{Einstein_EOM}) is of the form $A(z)=\sqrt{-V_{0}/3} \,z + cst$ (the constant can be eliminated by rescaling $x^{\alpha}$) and the metric reads
\beq
\label{g_DW_AdS}
ds^2 = e^{2 z/L} \, \eta_{\alpha \beta}   \, dx^{\alpha} dx^{\beta} \,\,+\,\, dz^2 \ ,
\eeq
where $L^2=-3/V_{0}$ is the AdS radius. By applying the radial coordinate redefinition \mbox{$r=L \, e^{-z/L}$}, the most familiar form of the AdS metric $\,ds^2 = \frac{L^2}{r^2} \, (\eta_{\alpha \beta}   \, dx^{\alpha} dx^{\beta} \,\,+\,\, dr^2 )\,$ is recast. The  AdS boundary ($z \rightarrow \infty$) is mapped to $r=0$ and the deep interior ($z \rightarrow -\infty$) to $r=\infty$.

\subsubsection{Glossary of AdS critical points at $\omega=0$}

\begin{table}[t!] 
\renewcommand{\arraystretch}{1.25}
\begin{center}
\scalebox{0.80}{
\begin{tabular}{|c|c|c|c|c|c|c|c|c|c|}
\hline 
 SUSY & $G_{0}$  & $g^{-2}\,V_{0}$ & $|W_{1}|$ & $|W_{\hat{1}}|$ &$\lambda_{0}$ & $\alpha_{0}$  & $\lambda'_{0}$  & $\phi_{0}$  &  Stability\\ 
\hline \hline
$\cN = 8$ & $\textrm{SO}(8)$ & $-6$  & $1$ & $1$ & $0$ & $0$ & $0$ & $0$ & $\checkmark$\\[2mm]
\hline
$\cN = 2$ & $\textrm{U}(3)$ & $-7.794$ & $1.140$  & $1.140$ & $0.275$  & $0$  & $0.329$ & $\pm\frac{\pi}{2}$  & $\checkmark$\\[2mm]
\hline
\multirow{4}{*}{$\cN = 1$} & \multirow{4}{*}{$\textrm{G}_2$} & \multirow{4}{*}{$-7.192$} & \multirow{2}{*}{$1.095^{*}$} & \multirow{2}{*}{$1.341$} & \multirow{4}{*}{$0.259$} & \multirow{4}{*}{$\pm 0.310 \pi$} & \multirow{4}{*}{$0.259$} & $\pm 0.310 \pi$  & \multirow{4}{*}{$\checkmark$}\\
\cline{9-9}
 &  &  &  &  &  &  & & $\pm 1.310 \pi$  & \\
 \cline{4-5}\cline{9-9}
 &  &  &  \multirow{2}{*}{$1.341$} & \multirow{2}{*}{$1.095^{*}$}  &  &   &   & $\mp 0.310 \pi$  & \\
 \cline{9-9}
 &  &  &  &  &  &  & & $\mp 1.310 \pi$  & \\[1mm]
\hline
\multirow{4}{*}{$\cN = 0$} & \multirow{4}{*}{$\textrm{SO}(7)_{\pm}$} & \multirow{2}{*}{$-6.687$}  & \multirow{2}{*}{$1.227$} & \multirow{2}{*}{$1.227$}  & \multirow{2}{*}{$0.201$} &  \multirow{2}{*}{$0$}  & \multirow{2}{*}{$0.201$} & $0$ & \multirow{4}{*}{$\times$}\\
\cline{9-9}
 &  &   & &  &  &  &  & $\pi$ & \\
\cline{3-9}
 &  & \multirow{2}{*}{$-6.988$} & \multirow{2}{*}{$1.254$} & \multirow{2}{*}{$1.254$}   & \multirow{2}{*}{$0.241$} & \multirow{2}{*}{$\pm\frac{\pi}{2}$}  & \multirow{2}{*}{$0.241$}  & $\pm\frac{\pi}{2}$ & \\
 \cline{9-9}
 &  &  & &  &  &   &  &  $\mp\frac{\pi}{2}$ & \\[1mm]
\hline
$\cN = 0$ & $\textrm{SU}(4)$ & $-8$  & $\frac{3}{2}$ & $\frac{3}{2}$ & $0$ & $0$ & $0.441$ & $\pm \frac{\pi}{2}$ & $\times$\\[1mm]
\hline
\end{tabular}}
\caption{{\it The $\textrm{SU}(3)$-invariant critical points of the $\textrm{SO}(8)$ gauged supergravity at $\omega=0$. For those solutions preserving $\mathcal{N}=1$, the mark $^{*}$ singles out the superpotential $( W_{1} \textrm{ vs } W_{\hat{1}})$ with respect to which supersymmetry is preserved. }}
\label{Table:c=0_points} 
\end{center}
%\vspace{-0.5cm}
\end{table}

The structure of SU(3)-invariant critical points of the purely electric SO(8) gauged supergravity at $\,\omega=0\,$ was classified thirty years ago by Warner in ref.~\cite{Warner:1983vz}. In this case, the theory is known to contain an  AdS solution at the origin preserving $\,\mathcal{N}=8\,$ supersymmetry and $G_{0}=\textrm{SO}(8)$ residual symmetry as well as other five types of AdS critical points preserving smaller (super)symmetry. The relevant data for these points\footnote{An exact form is known for the numbers in Table~\ref{Table:c=0_points} (see appendix~A in ref.~\cite{Bobev:2010ib}). We typed the numerical values in order to compare with other tables in the text for which only numerical values are available.} is summarised in Table~\ref{Table:c=0_points}.

\begin{table}[t!] 
\renewcommand{\arraystretch}{1.25}
\begin{center}
\scalebox{0.80}{
\begin{tabular}{|c|c|c|c|c|c|c|c|c|c|}
\hline 
 SUSY & $G_{0}$  & $g^{-2}\,V_{0}$ & $|W_{1}|$ & $|W_{\hat{1}}|$ &$\lambda_{0}$ & $\alpha_{0}$  & $\lambda'_{0}$  & $\phi_{0}$  &  Stability\\ 
\hline \hline
$\cN = 8$ & $\textrm{SO}(8)$ & $-6$  & $1$ & $1$ & $0$ & $0$ & $0$ & $0$ & $\checkmark$\\[2mm]
\hline
\multirow{3}{*}{$\cN = 2$} & \multirow{3}{*}{$\textrm{U}(3)$} & \multirow{3}{*}{$-8.354$} & \multirow{3}{*}{$1.180$}  & \multirow{3}{*}{$1.180$} & \multirow{3}{*}{$0.315$}  & $0.171 \pi$  & \multirow{3}{*}{$0.375$} & $\pm \frac{\pi}{2}$  & \multirow{3}{*}{$\checkmark$} \\
\cline{7-7}\cline{9-9}
&  &   &   &   &   & \multirow{2}{*}{$1.329 \pi$}  &   &  $0$ &  \\
\cline{9-9}
&  &   &   &   &   &   &   & $\pi$  &  \\
\hline
\multirow{8}{*}{$\cN = 1$} & \multirow{8}{*}{$\textrm{G}_2$} & \multirow{8}{*}{$-7.943$} & \multirow{4}{*}{$1.151^{*}$} & \multirow{4}{*}{$1.409$} & \multirow{8}{*}{$0.329$} & \multirow{2}{*}{$0.373 \pi$} & \multirow{8}{*}{$0.329$} & $ 0.373 \pi$  & \multirow{8}{*}{$\checkmark$}\\
 \cline{9-9}
 &  &  &  &  &  &  & & $1.373 \pi$  & \\
\cline{7-7}\cline{9-9}
 &  &  &  &  &  & \multirow{2}{*}{$1.127 \pi$} & & $1.127 \pi$  & \\
 \cline{9-9}
 &  &  &  &  &  &  & & $0.127 \pi$  & \\
\cline{4-5}\cline{7-7}\cline{9-9}
 &  &  &  \multirow{4}{*}{$1.409$} & \multirow{4}{*}{$1.151^{*}$}  &  &  \multirow{2}{*}{$0.373 \pi$} &   & $-0.373 \pi$  & \\
  \cline{9-9}
 &  &  &  &  &  &  & & $-1.373 \pi$  & \\
 \cline{7-7}\cline{9-9}
 &  &  &  &  &  & \multirow{2}{*}{$1.127 \pi$} & & $-1.127 \pi$  & \\
 \cline{9-9}
 &  &  &  &  &  &  & & $-0.127 \pi$  & \\
\hline
\multirow{6}{*}{$\cN = 0$} & \multirow{6}{*}{$\textrm{SO}(7)_{\pm}$} & \multirow{3}{*}{$-6.748$}  & \multirow{3}{*}{$1.232$} & \multirow{3}{*}{$1.232$}  & \multirow{3}{*}{$0.210$} &  \multirow{2}{*}{$0$}  & \multirow{3}{*}{$0.210$} & $0$ & \multirow{6}{*}{$\times$}\\
\cline{9-9}
 &  &   & &  &  &  &  & $\pi$ & \\
 \cline{7-7}\cline{9-9}
  &  &   & &  &  & $-\frac{\pi}{2}$ &  & $\pm\frac{\pi}{2}$ & \\
  \cline{3-9}
 &  & \multirow{3}{*}{$-7.771$} & \multirow{3}{*}{$1.322$} & \multirow{3}{*}{$1.322$}   & \multirow{3}{*}{$0.320$} & \multirow{2}{*}{$\pi$}  & \multirow{3}{*}{$0.320$}  & $0$ & \\
\cline{9-9}
 &  &   & &  &  &  &  & $\pi$ & \\
 \cline{7-7}\cline{9-9}
  &  &   & &  &  & $\frac{\pi}{2}$ &  & $\pm\frac{\pi}{2}$ & \\[1mm]
  \hline
\multirow{3}{*}{$\cN = 0$} & \multirow{3}{*}{$\textrm{SU}(4)$} & \multirow{3}{*}{$-8.581$}  & \multirow{3}{*}{$1.553$} & \multirow{3}{*}{$1.553$} & \multirow{3}{*}{$0.115$} & \multirow{2}{*}{$\pi$} & \multirow{3}{*}{$0.488$} & $0$ & \multirow{3}{*}{$\times$} \\
\cline{9-9}
 &  &   & &  &  &  &  & $\pi$ &  \\
\cline{7-7}\cline{9-9}
  &  &   & &  &  & $\frac{\pi}{2}$ &  & $\pm\frac{\pi}{2}$ &  \\[1mm]
\hline
\end{tabular}}
\caption{{\it The shifted $\textrm{SU}(3)$-invariant critical points of the $\textrm{SO}(8)$ gauged supergravity at $\omega=\frac{\pi}{8}$. These points have a counterpart at $\,\omega=0$. For those solutions preserving $\mathcal{N}=1$, the mark $^{*}$ singles out the superpotential $( W_{1} \textrm{ vs } W_{\hat{1}})$ with respect to which supersymmetry is preserved. }}
\label{Table:c=sqrt2-1_points}
\end{center}
\vspace{-0.5cm}
\end{table}

On the other hand, the issues of perturbative stability and higher-dimensional origin of these critical points have also been thoroughly investigated (see ref.~\cite{Nicolai:2011cy} for a list of references).
%
%\begin{itemize}
%
%\item[$\circ$] \textit{$\mathcal{N}=8$ , $\textrm{G}_{0}=\textrm{SO}(8)\,$ point :} This AdS critical point located at $\Sigma^{i}=0$ preserves the largest possible amount of (super)symmetry, so all the scalars of maximal supergravity get a mass at the conformal value $m^2 L^2 = -2$. Its lifting to 11d supergravity was worked out in ref.~\cite{Freund:1980xh}.
%
%\item[$\circ$] \textit{$\mathcal{N}=2$ , $\textrm{G}_{0}=\textrm{U}(3)\,$ point :} The scalar mass spectrum at this critical point can be found in ref.~\cite{Nicolai:1985hs,Bobev:2010ib}, whereas its lifting to 11d was derived in ref.~\cite{Corrado:2001nv}.
%
%\item[$\circ$] \textit{$\mathcal{N}=1$ , $\textrm{G}_{0}=\textrm{G}_{2}\,$ points :} The full scalar spectrum for these critical points was obtained in ref.~\cite{Bobev:2010ib}. The embedding into 11d supergravity was found earlier in ref.~\cite{deWit:1984nz}.
%
%\item[$\circ$] \textit{$\mathcal{N}=0$ , $\textrm{G}_{0}=\textrm{SO}(7)_{\pm}\,$ points :} The higher-dimensional origin of these points have been well explored in the literature \cite{Englert:1982vs,deWit:1983gs,deWit:1984nz,deWit:1984va}. The spectrum of scalar fluctuations contains tachyons below the B.F. bound, hence rendering the solutions unstable \cite{deWit:1983gs}.
%
%\item[$\circ$] \textit{$\mathcal{N}=0$ , $\textrm{G}_{0}=\textrm{SU}(4)\,$ point :} This solution was embedded into 11d supergravity in ref.~\cite{Pope:1984bd} and more recently found to be unstable under scalar fluctuations in ref.~\cite{Bobev:2010ib}.
%
%\end{itemize}
%
The analysis of the SU(3)-invariant sector at $\omega=0$ showed that, whenever supersymmetry did not protect solutions to have instabilities, these showed up somewhere in the full ${\mathcal{N}=8}$ spectrum. However, counterexamples to this were found soon after by analysing the SO(4)-invariant sector of the theory still with $\omega=0$ \cite{Warner:1983du,Fischbacher:2010ec} as well as within the G$_{2}$-invariant sector with $\omega\neq 0$ \cite{Borghese:2012qm}. The scalar mass spectra at these points turned out \cite{Borghese:2012qm,Borghese:2013dja} to be independent of $\omega$.

\subsubsection{Glossary of critical points at $\omega \neq 0$}

\begin{table}[t!] 
\renewcommand{\arraystretch}{1.25}
\begin{center}
\scalebox{0.80}{
\begin{tabular}{|c|c|c|c|c|c|c|c|c|c|}
\hline 
 SUSY & $G_{0}$  & $g^{-2}\,V_{0}$ & $|W_{1}|$ & $|W_{\hat{1}}|$ &$\lambda_{0}$ & $\alpha_{0}$  & $\lambda'_{0}$  & $\phi_{0}$  &  Stability\\ 
\hline \hline
\multirow{4}{*}{$\cN = 1$} & \multirow{4}{*}{$\textrm{G}_{2}$} & \multirow{4}{*}{$-7.040$}  & \multirow{2}{*}{$1.083^{*}$} & \multirow{2}{*}{$1.327$} & \multirow{4}{*}{$0.242$} & \multirow{4}{*}{$-\frac{\pi}{4}$} & \multirow{4}{*}{$0.242$} & $-\frac{\pi}{4}$ & \multirow{4}{*}{$\checkmark$} \\
\cline{9-9}
 &  &   & &  &  &  &  & $\frac{3 \pi}{4}$ &  \\
 \cline{4-5}\cline{9-9}
  &  &   & \multirow{2}{*}{$1.327$} & \multirow{2}{*}{$1.083^{*}$} &  &  &  & $\frac{\pi}{4}$ & \\
\cline{9-9}
 &  &   & &  &  &  &  & $-\frac{3 \pi}{4}$ &  \\
\hline
\multirow{4}{*}{$\cN = 1$} & \multirow{4}{*}{$\textrm{SU}(3)$} & \multirow{4}{*}{$-10.392$}  & \multirow{2}{*}{$1.316^*$} & \multirow{2}{*}{$2.632$} & \multirow{4}{*}{$0.275$} & \multirow{4}{*}{$\frac{3 \pi}{4}$} & \multirow{4}{*}{$0.573$} & $\frac{\pi}{4}$ & \multirow{4}{*}{$\checkmark$} \\
\cline{9-9}
 &  &   &  &  &  &  &  & $-\frac{3 \pi}{4}$ & \\
\cline{4-5}\cline{9-9}
 &  &   & \multirow{2}{*}{$2.632$} & \multirow{2}{*}{$1.316^*$} &  &  &  & $-\frac{\pi}{4}$ & \\
\cline{9-9}
  &  &   &  &  &  &  &  & $\frac{3 \pi}{4}$ & \\ [1mm]
\hline
\multirow{4}{*}{$\cN = 0$} & \multirow{4}{*}{$\textrm{G}_{2}$} & \multirow{4}{*}{$-10.170$}  & \multirow{2}{*}{$2.762$} & \multirow{2}{*}{$1.595$} & \multirow{2}{*}{$0.467$} & \multirow{2}{*}{$\frac{3\pi}{4}$} & \multirow{2}{*}{$0.467$} & $\frac{3\pi}{4}$ & \multirow{4}{*}{$\checkmark$} \\
\cline{9-9}
 &  &   & &  &  &  &  & $-\frac{\pi}{4}$ &  \\
 \cline{4-9}
  &  &   & \multirow{2}{*}{$1.595$} & \multirow{2}{*}{$2.762$} & \multirow{2}{*}{$0.467$} & \multirow{2}{*}{$\frac{3\pi}{4}$} & \multirow{2}{*}{$0.467$} & $-\frac{3\pi}{4}$ & \\
\cline{9-9}
 &  &   & &  &  &  &  & $\frac{\pi}{4}$ &  \\
\hline
\multirow{8}{*}{$\cN = 0$} & \multirow{8}{*}{$\textrm{SU}(3)$} & \multirow{8}{*}{$-10.237$} & \multirow{4}{*}{$2.747$} & \multirow{4}{*}{$1.467$} & \multirow{8}{*}{$0.400$} & \multirow{2}{*}{$0.702 \pi$} & \multirow{8}{*}{$0.512$} & $ 0.785 \pi$  & \multirow{8}{*}{$\checkmark$ [ see (\ref{new_SU3_spectrum}) ]   }\\
 \cline{9-9}
 &  &  &  &  &  &  & & $1.785 \pi$  & \\
\cline{7-7}\cline{9-9}
 &  &  &  &  &  & \multirow{2}{*}{$0.798 \pi$} & & $-0.285 \pi$  & \\
 \cline{9-9}
 &  &  &  &  &  &  & & $-1.285 \pi$  & \\
\cline{4-5}\cline{7-7}\cline{9-9}
 &  &  &  \multirow{4}{*}{$1.467$} & \multirow{4}{*}{$2.747$}  &  &  \multirow{2}{*}{$0.702 \pi$} &   & $-0.785 \pi$  & \\
  \cline{9-9}
 &  &  &  &  &  &  & & $-1.785 \pi$  & \\
 \cline{7-7}\cline{9-9}
 &  &  &  &  &  & \multirow{2}{*}{$0.798 \pi$} & & $0.285 \pi$  & \\
 \cline{9-9}
 &  &  &  &  &  &  & & $1.285 \pi$  & \\
\hline
\end{tabular}}
\caption{{\it The genuine $\textrm{SU}(3)$-invariant critical points of the $\textrm{SO}(8)$ gauged supergravity at $\omega=\frac{\pi}{8}$. These points have no counterpart at $\,\omega=0$. For those solutions preserving $\mathcal{N}=1$, the mark $^{*}$ singles out the superpotential $( W_{1} \textrm{ vs } W_{\hat{1}})$ with respect to which supersymmetry is preserved. }}
\label{Table:c=sqrt2-1_new_points} 
\end{center}
\vspace{-0.5cm}
\end{table}

The structure of SU(3)-invariant critical points at $\omega \neq 0$ was explored in ref.~\cite{Borghese:2012zs} using a superpotential differing from (\ref{W1z-zeta}) by an overall phase, as discussed in section~\ref{sec:superpotentials}. However it is clear from (\ref{V_from_W}) that the scalar potential is not sensitive to overall phases, so the critical points associated to (\ref{W1z-zeta}) coincide with those found in ref.~\cite{Borghese:2012zs}. Turning on $\omega$ was found to modify the location and energy of the critical points existing at $\omega=0$ as well as to create new ones with no counterpart at $\omega=0$. As a check of the scalar potential in (\ref{VSU3}), we have exhaustively verified the set of critical points found in ref.~\cite{Borghese:2012zs}. The entire set of AdS solutions can be divided into two \mbox{categories} : 
\begin{itemize}

\item[$i)$] points which are \textit{shifted counterparts} of those at $\omega=0$ : These points have the \textit{same} normalised mass spectra as their counterparts at $\omega=0$, hence inheriting their stability properties \cite{Dall'Agata:2012bb,Borghese:2012qm,Borghese:2012zs}. The list of these points at $\,\omega=\frac{\pi}{8}\,$ is shown in Table~\ref{Table:c=sqrt2-1_points}.

\item[$ii)$] points with \textit{no counterpart} at $\omega=0$ : These points are genuinely associated to \textit{dyonic} SO(8) gaugings. There are novel ${\mathcal{N}=1}$ AdS solutions with either $\textrm{G}_{2}$ or $\textrm{SU}(3)$ residual symmetry as well as non-supersymmetric critical points preserving  either $\textrm{G}_{2}$ or $\textrm{SU}(3)$ too.  The set of these points at $\omega=\frac{\pi}{8}$ is summarised in Table~\ref{Table:c=sqrt2-1_new_points}.

\end{itemize}
 
Perturbative stability of the non-supersymmetric point preserving $\textrm{G}_{2}$ was checked in refs~\cite{Borghese:2012qm,Borghese:2012zs}. However, the lack of a derivation from scratch of the \mbox{$\omega$-dependent} supergravity quantities, concretely of the fermi mass terms, made an analysis of stability for the novel non-supersymmetric and SU(3)-preserving point impossible. Now we are at the position to perform such an analysis here.

\subsubsection{Stability of the new $\mathcal{N}=0$ , $\textrm{G}_{0}=\textrm{SU}(3)$ critical point}

This AdS solution was shown to have $\omega$-dependent mass spectra in ref.~\cite{Borghese:2012zs}. Furthermore, the scalar masses for the SU(3)-singlets were computed at $\omega=\frac{\pi}{8}$ and found to satisfy the B.F. bound, but the stability of the full scalar spectrum remained an open question. Plugging the VEVs of the scalars displayed in Table~\ref{Table:c=sqrt2-1_new_points} into the fermion mass terms derived in the previous section, it is now straightforward to compute the full scalar mass spectrum via the mass formula in (\ref{Mass-matrix_scalars}). The outcome for the 70 scalar masses at $\,\omega=\frac{\pi}{8}\,$ is

\beq
\label{new_SU3_spectrum}
\begin{array}{ccrrrrrrrr}
m^2 \, L^2 &=& 6.223 \,\,(\times 1) & , &  5.914 \,\,(\times 1) & , & 1.138 \,\,(\times 1) & , & -1.275 \,\,(\times 1) & , \\
& & -1.641 \,\,(\times 12) & , & -0.908 \,\,(\times 12) & , & -1.504 \,\,(\times 8) & , & -0.235 \,\,(\times 8) & , \\
& & -1.073 \,\,(\times 6) & , & 0 \,\,(\times 20) & ,
\end{array}
\eeq
where the four masses in the first row correspond to the SU(3)-singlets in the truncated theory. The rest of the scalar masses, however, could not be computed before and show that this point is \textit{perturbatively stable} with respect to fluctuations of \textit{all} the scalars in maximal supergravity.  As usual in supergravity theories, the masses of some of the non-singlet fields are smaller than those of the singlets. Notice also the presence of $20$ massless fields (Goldstone bosons) reflecting the spontaneous $\,\textrm{SO}(8) \rightarrow \textrm{SU}(3)\,$ symmetry breaking at this vacuum.

The masses for the vectors after the symmetry breaking can also be computed immediately using again the fermi mass terms we derived and the mass formula in (\ref{Mass-matrix_vectors}). At $\,\omega=\frac{\pi}{8}\,$, they are given by
\beq
\label{new_SU3_spectrum_vectors}
\begin{array}{ccrrrrrrrr}
m^2 \, L^2 &=& 4.520 \,\,(\times 1) & , &  2.321 \,\,(\times 1) & , & 3.194 \,\,(\times 6) & , & 2.757 \,\,(\times 6) & , \\
& & 0.128 \,\,(\times 6) & , &  0 \,\,(\times 8) & ,
\end{array}
\eeq
where one identifies the eight massless vectors associated to the SU(3) residual symmetry. The first two masses correspond to the SU(3)-singlets and reflect the complete breaking of the $\textrm{U}(1) \times \textrm{U}(1)$ gauging in the truncated theory. 

We want to highlight that (up to our knowledge) this is the first example of a non-supersymmetric and nevertheless \textit{fully stable} critical point in new maximal supergravity with a scalar mass spectrum being \textit{sensitive} to the electromagnetic phase $\omega$. Previous stable cases were insensitive \cite{Dall'Agata:2012bb,Borghese:2012qm} and those being sensitive were unstable \cite{Dall'Agata:2012sx,Borghese:2013dja}.

\subsection{BPS domain-walls}

We now move to study BPS domain-wall configurations where the scalars develop a profile $\Sigma^{i}(z)$ and the scale factor $A(z)$ is no longer linear in $z$. By plugging the domain-wall Ansatz (\ref{g_DW}) into the action (\ref{S_scalar}) one finds
\beq
\label{S_density}
S_{\textrm{DW}}(A,\Sigma^{i}) = \frac{a}{2} \, \int_{-\infty}^{\infty} dz \,  e^{3 A} \left[ \, 6\, (\partial_{z} A)^2  -  K_{ij} (\partial_{z} \Sigma^{i})  (\partial_{z} \Sigma^{j}) - 2\, V(\Sigma^{i}) \, \right] \ ,
\eeq
where $\,a\,$ is the area transverse to the domain-wall direction. The energy per unit of transverse area is then given by \cite{Skenderis:1999mm} 
\beq
\label{E_density}
\begin{array}{ccl}
E_{\textrm{DW}}(A,\Sigma^{i}) &=& -\frac{1}{a} \, S_{\textrm{DW}}(A,\Sigma^{i}) \\[1mm]
& = & \frac{1}{2} \, \int_{-\infty}^{\infty} dz \,  e^{3 A} \left[ \, -6\, (\partial_{z} A)^2  +  K_{ij} (\partial_{z} \Sigma^{i})  (\partial_{z} \Sigma^{j}) + 2\, V(\Sigma^{i}) \, \right] \ .
\end{array}
\eeq
The fact that $V$ can be obtained from a superpotential as (\ref{V_from_W}) allows the energy density (\ref{E_density}) to be written \`a la Bogomol'nyi (completing squares) by using the relations in (\ref{constraintsW}) \cite{Ahn:2000mf}. Then it is extremised by BPS domain-wall solutions for which gravitational stability is guaranteed \cite{Skenderis:1999mm}. These solutions are found to satisfy the first-order set of equations\footnote{As shown in ref.~\cite{Ahn:2009as}, the gauge choice $\theta=\psi=0$ holds along the flow such that the kinetic function $T(\lambda',\phi,\theta,\psi)$ in (\ref{T_kinetic}) does not contribute to the energy density.}
\beq
\label{BPS_equations}
\begin{array}{cclcccccccc}
\partial_{z} A &=& \mp \, \sqrt{2} \, g \,  |W| & , \\[3mm]
\partial_{z} \lambda  &=& \pm  \, g \, \dfrac{\sqrt{2}}{3} \, \partial_{\lambda} |W| & \hspace{3mm} , \hspace{3mm} & \partial_{z} \alpha  &=& \pm  \, g \, \dfrac{4\sqrt{2}}{3 \sinh^2(2 \lambda) } \, \partial_{\alpha} |W| & , \\[4mm]
\partial_{z} \lambda'  &=& \pm  \, g \, \dfrac{1}{2 \sqrt{2}} \, \partial_{\lambda'} |W| & \hspace{3mm} , \hspace{3mm} & \partial_{z} \phi  &=& \pm  \, g \, \dfrac{\sqrt{2}}{\sinh^2(2 \lambda') } \, \partial_{\phi} |W| & ,
\end{array}
\eeq
and connect two supersymmetric AdS points at $z=\pm \infty$ along a steepest descent path\footnote{The actual flow occurs in the opposite direction as $V|_{z\rightarrow \pm\infty} \sim -6 g^2|W|^2$ and it runs from higher to lower values of the potential.} of $|W|$.  At the two end points, one has $\partial_{z}\Sigma^{i} \propto \partial_{\Sigma^{i}}|W| =0$ and, using the AdS/CFT correspondence,  the dual field theory is conjectured to flow from an UV fixed point at the boundary of AdS ($z\rightarrow +\infty$) to an IR fixed point at the deep interior ($z\rightarrow -\infty$). When approaching these asymptotic regions, the scale factor behaves as $\,\left. A(z) \sim L^{-1}  z\,\right|_{z\rightarrow \pm \infty}\,$ with gradients $\,L^{-1}_{\textrm{UV}}=\mp g \, \sqrt{2} \, |W_{\textrm{UV}}|\,$ and $\,L^{-1}_{\textrm{IR}}=\mp g \, \sqrt{2} \, |W_{\textrm{IR}}|\,$, respectively.

Using the (inverse) metric $K_{ij}$ in (\ref{K-metric}), the flow equations in (\ref{BPS_equations}) can be written in the more compact form
\beq
\label{BPS-equations-compact}
\partial_{z} A = \mp \, \sqrt{2} \, g \,  |W|
\hspace{10mm} \textrm{ and } \hspace{10mm}
\partial_{z} \Sigma^{i}  = \pm  2 \, \sqrt{2} \, g \,  K^{ij} \, \partial_{\Sigma^{j}} |W|\ .
\eeq
Near the asymptotic regions at $z\rightarrow \pm \infty$, the non-linear flow is well approximated by a linear one satisfying
\beq
\label{Delta_matrix}
\partial_{z} \Sigma^{i}  \sim -\frac{1}{L_{0}} \, {\Delta^{i}}_{j} \, (\Sigma^{j} - \Sigma_{0}^{j} ) \ ,
\eeq
where $\Sigma^{i}_{0}$ denote the VEVs of the scalars at one of the asymptotic AdS points, $L_{0}$ is the AdS radius and the matrix ${\Delta^{i}}_{j}$ is also evaluated at that point. The eigenvalues of ${\Delta^{i}}_{j}$ encode the masses of the fields and therefore also the conformal dimension of the dual operators.

The aim is to solve the set  (\ref{BPS-equations-compact}) of differential equations numerically using the $\omega$-dependent superpotential $\,|W|\,$ in section~\ref{sec:superpotentials}. From now on, we will take $\,W=W_{1}\,$ in (\ref{W1&What1}) without entailing a loss of generality\footnote{To be consistent with this choice, one has to select the ($*$-marked) AdS solutions preserving $W_{1}$ and \textit{not} $W_{\hat{1}}$ for those domain-walls flowing towards $\mathcal{N}=1$ points in Tables~\ref{Table:c=sqrt2-1_points} and \ref{Table:c=sqrt2-1_new_points}.} and investigate two types of BPS domain-walls: $\,i)$ domain-walls flowing between two supersymmetric points in Table~\ref{Table:c=sqrt2-1_points}, so they can be understood as $\omega$-deformations of others existing at $\omega=0$  $\,\,\,ii)$ domain-walls which have no counterpart at $\omega=0$ as they flow towards some of the genuine supersymmetric points in Table~\ref{Table:c=sqrt2-1_new_points}.

\subsubsection{Domain-walls with $\omega=0$ counterpart}

Examples of BPS domain-walls interpolating between the ${\mathcal{N}=8}\,\&\,\textrm{SO}(8)$ point and either the \mbox{${\mathcal{N}=1}\,\&\,\textrm{G}_{2}$} or the ${\mathcal{N}=2}\,\&\,{\textrm{U}(3)}$ point have been constructed in the electric case of $\omega=0$ \cite{Ahn:2000aq,Ahn:2001kw,Bobev:2009ms}. In addition, the connection to BLG theory and deformations thereof by adding mass terms was put forward in refs~\cite{Ahn:2008ya,Ahn:2008gda,Bobev:2009ms}.  Here we will numerically solve the first-order equations in (\ref{BPS-equations-compact}) to determine how the flows get modified when turning on the electromagnetic phase $\omega$. 

\begin{figure}[t!]
\begin{center}
\includegraphics[width=65mm]{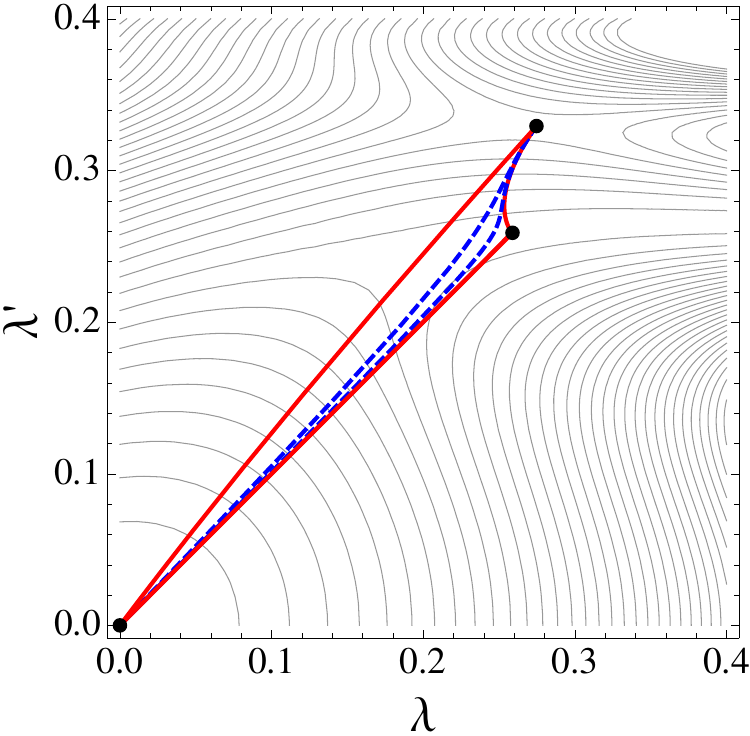}
\hspace{2mm}
\includegraphics[scale=0.74,trim = 0mm -10mm 0mm 0mm]{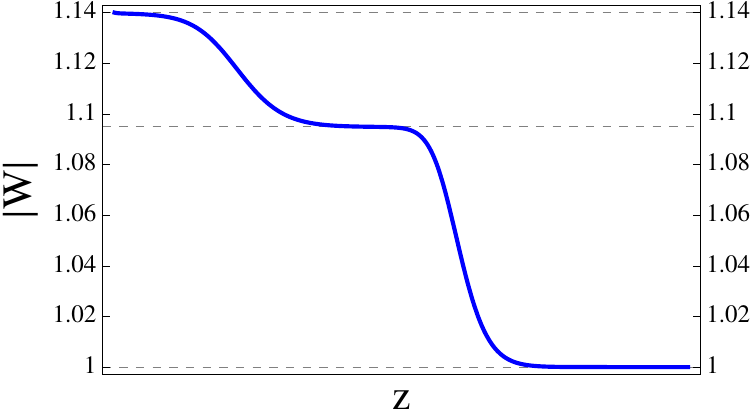}
\caption{{\it Contours of the superpotential at $\omega=0$ showing BPS domain-walls interpolating between $\textrm{SO}(8) \leftrightarrow \textrm{U}(3)$ (upper straight red line),  $\textrm{SO}(8) \leftrightarrow \textrm{G}_{2}$ (lower straight red line) and $\textrm{G}_{2} \leftrightarrow \textrm{U}(3)$ (curved red line) supersymmetric AdS vacua. The dashed blue lines represent generic $\textrm{SO}(8) \leftrightarrow \textrm{U}(3)$ steepest descents passing arbitrarily close to the $\textrm{G}_{2}$ point. One of these arbitrarily close paths is depicted in the right plot.}}
\label{Fig:Old_DW_Omega0} 
\end{center}
\vspace{-5mm}
\end{figure}

In order to plot the flows of the four-field superpotential $W(\lambda,\alpha,\lambda',\phi)$, it is necessary to take a two-dimensional slice. We will take
\beq
W(\lambda,\lambda') = W(\lambda,\alpha^{*},\lambda',\phi^{*}) \ ,
\eeq
with sections $\,\alpha^{*}(\lambda')\,$ and $\,\phi^{*}(\lambda')\,$ of the form
\beq
\alpha^{*}=\frac{{\lambda'}^{2}-{\lambda'}_{(2)}^{2}}{{\lambda'}_{(1)}^{2}-{\lambda'}_{(2)}^{2}} \, \alpha_{(1)} + \frac{{\lambda'}^{2}-{\lambda'}_{(1)}^{2}}{{\lambda'}_{(2)}^{2}-{\lambda'}_{(1)}^{2}} \, \alpha_{(2)}
\hspace{4mm} , \hspace{4mm}
\phi^{*}=\frac{{\lambda'}^{2}-{\lambda'}_{(2)}^{2}}{{\lambda'}_{(1)}^{2}-{\lambda'}_{(2)}^{2}} \, \phi_{(1)} + \frac{{\lambda'}^{2}-{\lambda'}_{(1)}^{2}}{{\lambda'}_{(2)}^{2}-{\lambda'}_{(1)}^{2}} \, \phi_{(2)} \ .
\eeq
The above choice of slice is then guaranteed to catch pairs of critical points located at $\Sigma_{(1)}$ and $\Sigma_{(2)}$. Let us emphasise that, irrespective of the slicing, we are solving the actual system of first-order equations in (\ref{BPS-equations-compact}) and not any projected version of it.

\begin{figure}[t!]
\begin{center}
\includegraphics[width=65mm]{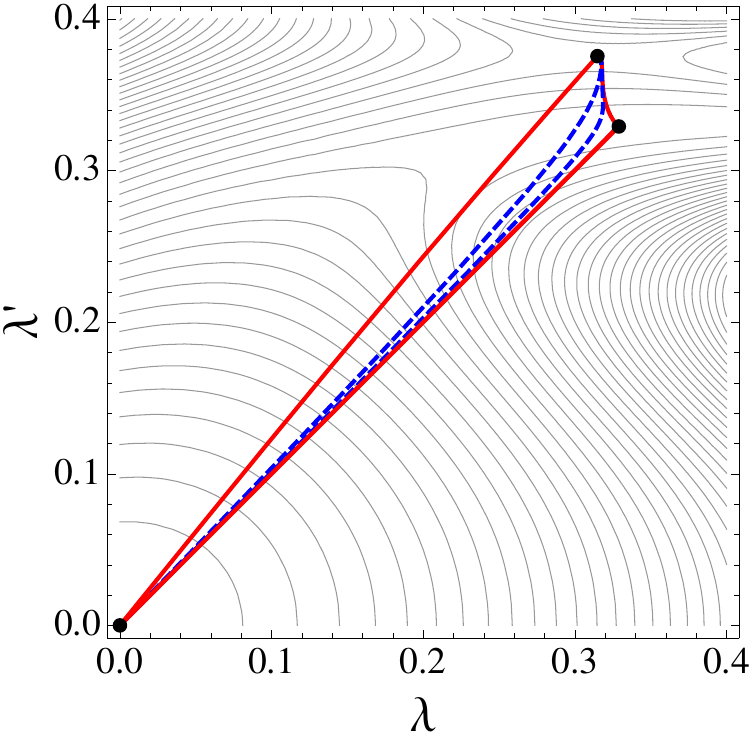}
\hspace{2mm}
\includegraphics[scale=0.74,trim = 0mm -10mm 0mm 0mm]{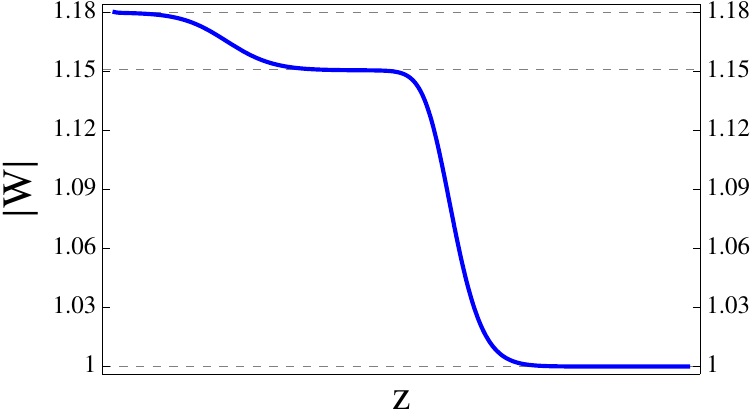}
\caption{{\it Contours of the superpotential at $\omega=\frac{\pi}{8}$ showing BPS domain-walls interpolating between $\textrm{SO}(8) \leftrightarrow \textrm{U}(3)$ (upper straight red line),  $\textrm{SO}(8) \leftrightarrow \textrm{G}_{2}$ (lower straight red line) and $\textrm{G}_{2} \leftrightarrow \textrm{U}(3)$ (curved red line) supersymmetric AdS vacua. The dashed blue lines represent generic $\textrm{SO}(8) \leftrightarrow \textrm{U}(3)$ steepest descents passing arbitrarily close to the $\textrm{G}_{2}$ point. One of these arbitrarily close paths is depicted in the right plot.}}
\label{Fig:Old_DW_OmegaPi8} 
\end{center}
\vspace{-5mm}
\end{figure}

Setting $\,\omega=0\,$, the ${\mathcal{N}=8}\,\&\,\textrm{SO}(8)$ point is located at the origin $\lambda_{0}=\lambda'_{0}=0$ whereas the other two points are located at
\beq
\begin{array}{lccccc}
{\mathcal{N}=1}\,\&\,\textrm{G}_{2} & \hspace{2mm} : \hspace{5mm} & \lambda_{0}=\lambda'_{0}=0.259  & , & \alpha_{0}=\phi_{0}=0.310 \pi  & , \\[1mm]
{\mathcal{N}=2}\,\&\,\textrm{U}(3)  & \hspace{2mm} : \hspace{5mm} & \lambda_{0}=0.275 \,\, , \,\, \lambda'_{0}=0.329  & , & \alpha_{0}=0 \,\, , \,\, \phi_{0}= \frac{\pi}{2}  &  ,
\end{array}
\eeq 
and correspond to values $|W_{0}|_{\textrm{SO}(8)}={1}$, $|W_{0}|_{\textrm{G}_{2}}={1.095}$ and $|W_{0}|_{\textrm{U}(3)}={1.140}$ of the superpotential. Generic solutions to the flow-equations typically run off to infinity but, as observed in ref.~\cite{Bobev:2009ms}, there is a (one parameter family) cone of flows from the ${\mathcal{N}=8}\,\&\,\textrm{SO}(8)$ point to the ${\mathcal{N}=2}\,\&\,\textrm{U}(3)$ point passing arbitrarily close to the ${\mathcal{N}=1}\,\&\,\textrm{G}_{2}$ point. This behaviour is illustrated in Figure~\ref{Fig:Old_DW_Omega0}. The scalar masses and the set of eigenvalues of ${\Delta^{i}}_{j}$ in (\ref{Delta_matrix}) read
\beq
\label{masses_Delta_G2}
\begin{array}{lccccc}
{\mathcal{N}=1}\,\&\,\textrm{G}_{2} & \hspace{2mm} : \hspace{5mm} & m^2 L^2 &=&   6.449 \, , \, -2.242  \,\, , \,\, 1.551 \,\, ,\,\,  -1.425 & , \\[1mm]
  & & \Delta &=& -1.449 \, , \,  1.408 \,\, , \,\, 3.449 \,\, ,\,\,  0.592& , \\[3mm]
{\mathcal{N}=2}\,\&\,\textrm{U}(3) & \hspace{2mm} : \hspace{5mm} & m^2 L^2 &=&   7.123 \, , \, 2.000  \,\, , \,\, 2.000 \,\, ,\,\,  -1.123 & , \\[1mm]
& &  \Delta &=& -1.562 \, , \,  3.562 \,\, , \,\, -0.562 \,\, ,\,\,  2.562&  .
\end{array}
\eeq 
Because of supersymmetry, the eigenvalues of ${\Delta^{i}}_{j}$ come in pairs adding to $2$. The \mbox{${\mathcal{N}=1}\,\&\,\textrm{G}_{2}$} point corresponds to one irrelevant operator of dimension $\,3-\Delta\,$ ($\Delta<0$), two non-normalisable modes ($0<\Delta< \frac{3}{2}$) and one normalisable mode ($\Delta>\frac{3}{2}$). The \mbox{${\mathcal{N}=2}\,\&\,\textrm{U}(3)$} point corresponds to two irrelevant operators of dimension \mbox{$\,3-\Delta\,$} ($\Delta<0$) and two normalisable modes ($\Delta>\frac{3}{2}$) \cite{Bobev:2009ms}.

Turning on $\,\omega=\frac{\pi}{8}\,$ shifts the location of the asymptotic AdS points and changes the profiles of the flows. Some steepest descents are depicted in Figure~\ref{Fig:Old_DW_OmegaPi8}.  There are again flows between the three AdS points located now at the origin and at 
\beq
\begin{array}{lccccc}
{\mathcal{N}=1}\,\&\,\textrm{G}_{2} & \hspace{1mm} : \hspace{1mm} & \lambda_{0}=\lambda'_{0}=0.329  & , & \alpha_{0}=\phi_{0}=0.373 \pi  & , \\[1mm]
{\mathcal{N}=2}\,\&\,\textrm{U}(3)  & \hspace{1mm} : \hspace{1mm} & \lambda_{0}=0.315 \,\, , \,\, \lambda'_{0}=0.375  & , & \alpha_{0}=0.171\pi \,\, , \,\, \phi_{0}= \frac{\pi}{2}  &  .
\end{array}
\eeq 
The corresponding values of the superpotential are $|W_{0}|_{\textrm{SO}(8)}={1}$, $|W_{0}|_{\textrm{G}_{2}}={1.151}$ and $|W_{0}|_{\textrm{U}(3)}={1.180}$. The scalar masses and the eigenvalues of ${\Delta^{i}}_{j}$ are not sensitive to the electromagnetic phase. Making a \textit{dyonic} choice of the gauging does not change the qualitative features of these flows. In particular, there is still a (one parameter family) cone of flows from the ${\mathcal{N}=8}\,\&\,\textrm{SO}(8)$ to the ${\mathcal{N}=2}\,\&\,\textrm{U}(3)$ point passing arbitrarily close to the ${\mathcal{N}=1}\,\&\,\textrm{G}_{2}$ point, as happened for their electric counterparts. However, as we will see next, a new flow connecting the ${\mathcal{N}=2}\,\&\,\textrm{U}(3)$ point to the novel\footnote{In order to avoid confusion between the ${\mathcal{N}=1}$ points preserving $\textrm{G}_{2}$ in Table~\ref{Table:c=sqrt2-1_points} and in Table~\ref{Table:c=sqrt2-1_new_points}, we have attached the labels $\textrm{G}_{2}$ and $\textrm{G}^{*}_{2}$ respectively.} ${\mathcal{N}=1}\,\&\,\textrm{G}^{*}_{2}$ point in Table~\ref{Table:c=sqrt2-1_new_points} also exists.

\subsubsection{Domain-walls without $\omega=0$ counterpart}

\begin{figure}[t!]
\begin{center}
\includegraphics[width=65mm]{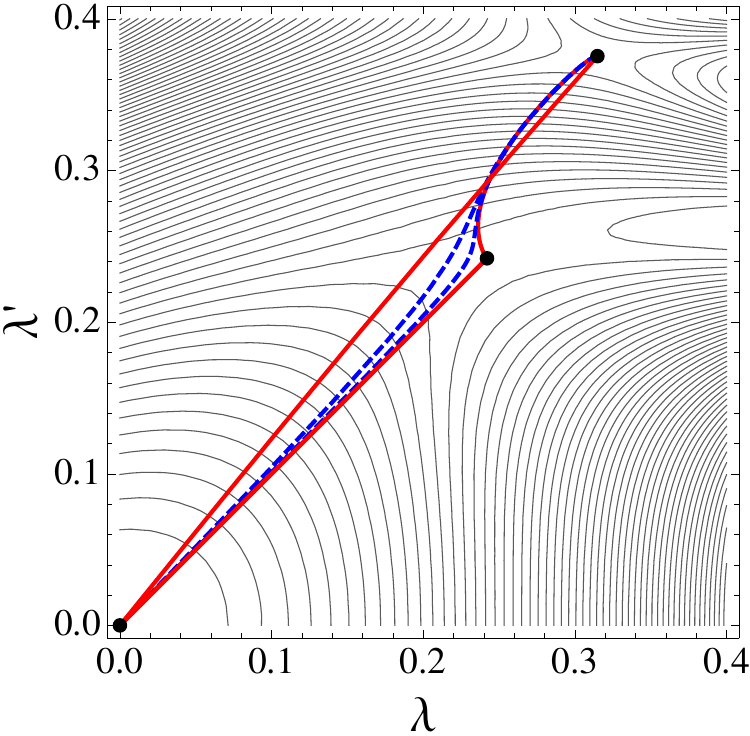}
\hspace{2mm}
\includegraphics[scale=0.74,trim = 0mm -10mm 0mm 0mm]{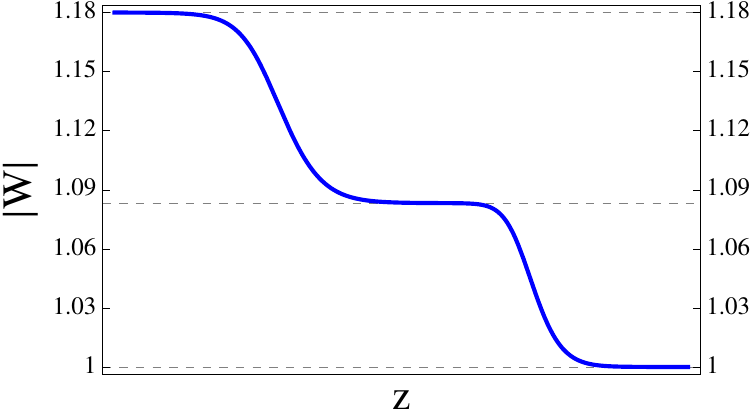}
\caption{{\it Contours of the superpotential at $\omega=\frac{\pi}{8}$ showing BPS domain-walls interpolating between $\textrm{SO}(8) \leftrightarrow \textrm{U}(3)$ (upper straight red line),  $\textrm{SO}(8) \leftrightarrow G^{*}_{2}$ (lower straight red line) and $G^{*}_{2} \leftrightarrow \textrm{U}(3)$ (curved red line) supersymmetric AdS vacua. The dashed blue lines represent generic $\textrm{SO}(8) \leftrightarrow \textrm{U}(3)$ steepest descents passing arbitrarily close to the $G^{*}_{2}$ point. One of these arbitrarily close paths is depicted in the right plot.}}
\label{Fig:NewG2_DW_OmegaPi8} 
\end{center}
\vspace{-5mm}
\end{figure}

Let us now describe \textit{dyonic} flows at $\omega=\frac{\pi}{8}$ which do not have an electric counterpart at $\omega=0$. These are flows involving either the novel ${\mathcal{N}=1}\,\&\,\textrm{G}^{*}_{2}$ or the ${\mathcal{N}=1}\,\&\,\textrm{SU}(3)$ points in Table~\ref{Table:c=sqrt2-1_new_points}.

\begin{figure}[t!]
\begin{center}
\includegraphics[width=65mm]{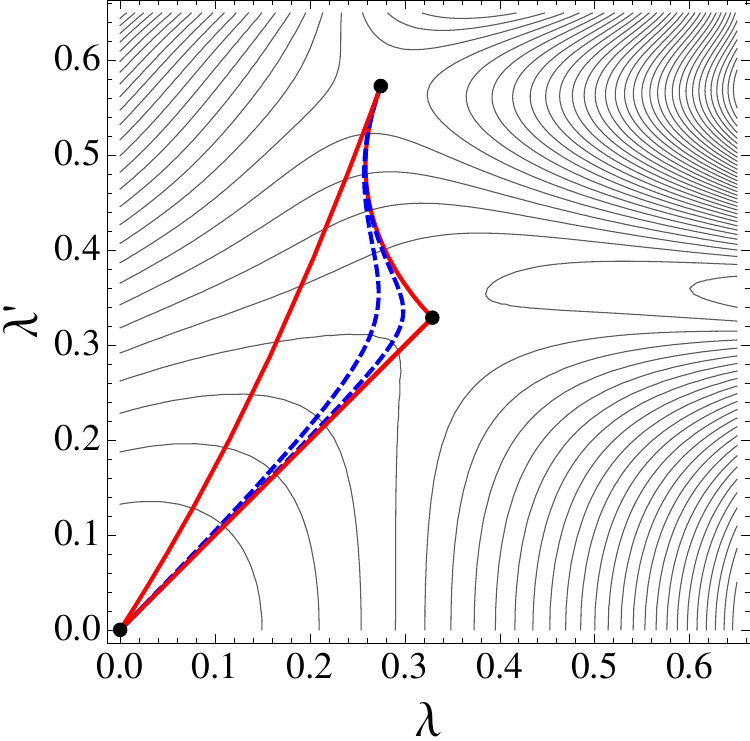}
\hspace{2mm}
\includegraphics[scale=0.74,trim = 0mm -10mm 0mm 0mm]{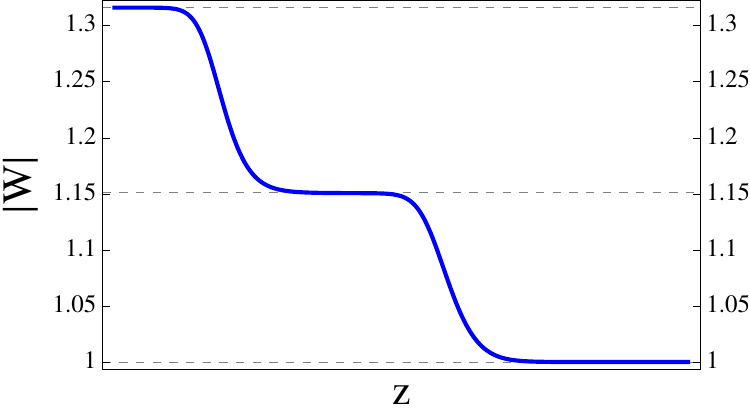}
\caption{{\it Contours of the superpotential at $\omega=\frac{\pi}{8}$ showing BPS domain-walls interpolating between $\textrm{SO}(8) \leftrightarrow \textrm{SU}(3) $ (upper almost straight red line),  $\textrm{SO}(8) \leftrightarrow \textrm{G}_{2}$ (lower straight red line) and $\textrm{G}_{2} \leftrightarrow \textrm{SU}(3)$ (curved red line) supersymmetric AdS vacua. The dashed blue lines represent generic $\textrm{SO}(8) \leftrightarrow \textrm{SU}(3)$ steepest descents passing arbitrarily close to the $\textrm{G}_{2}$ point. One of these arbitrarily close paths is depicted in the right plot.}}
\label{Fig:NewSU3_DW_OmegaPi8} 
\end{center}
\vspace{-5mm}
\end{figure}

In the first case, there are flows connecting the ${\mathcal{N}=8}\,\&\,\textrm{SO}(8)$ point at $\lambda_{0}=\lambda'_{0}=0$ to the genuine \mbox{${\mathcal{N}=1}\,\&\,\textrm{G}^{*}_{2}\,$} and the $\,{\mathcal{N}=2}\,\&\,\textrm{U}(3)$ points located at
\beq
\begin{array}{lccccc}
{\mathcal{N}=1}\,\&\,\textrm{G}^{*}_{2} & \hspace{1mm} : \hspace{1mm} & \lambda_{0}=\lambda'_{0}=0.242  & , & \alpha_{0}=-\frac{\pi}{4} \,\,,\,\, \phi_{0}=\frac{3\pi}{4}  & , \\[1mm]
{\mathcal{N}=2}\,\&\,\textrm{U}(3)  & \hspace{1mm} : \hspace{1mm} & \lambda_{0}=0.315 \,\, , \,\, \lambda'_{0}=0.375  & , & \alpha_{0}=0.171\pi \,\, , \,\, \phi_{0}= \frac{\pi}{2}  &  .
\end{array}
\eeq 
The superpotential takes values $|W_{0}|_{\textrm{SO}(8)}={1}$, $|W_{0}|_{\textrm{G}^{*}_{2}}={1.083}$ and $|W_{0}|_{\textrm{U}(3)}={1.180}$ at these points. In the linearised region around the ${\mathcal{N}=1}\,\&\,\textrm{G}^{*}_{2}\,$ point, the mass spectrum and the eigenvalues of ${\Delta^{i}}_{j}$ coincide with those for the ${\mathcal{N}=1}\,\&\,\textrm{G}_{2}$ point (\ref{masses_Delta_G2}). The behaviour of the steepest descents seems no longer as smooth as it was in Figure~\ref{Fig:Old_DW_OmegaPi8} even though a (one parameter family) cone of flows from the ${\mathcal{N}=8}\,\&\,\textrm{SO}(8)\,$ to the ${\mathcal{N}=2}\,\&\,\textrm{U}(3)\,$ point passing arbitrarily close to the ${\mathcal{N}=1}\,\&\,\textrm{G}^{*}_{2}\,$ point still exists. This time we observe paths, \textit{e.g.} the upper straight red line in Figure~\ref{Fig:NewG2_DW_OmegaPi8}, passing through these flows before getting the ${\mathcal{N}=2}\,\&\,\textrm{U}(3)$ point. This is a consequence of the choice of field variables we have used to build the steepest descents. We have verified this by applying the field redefinitions in (\ref{field-redef}). In terms of the new variables $(z,\zeta_{12})$, the set of red line paths displayed in the Figures \ref{Fig:Old_DW_OmegaPi8} and \ref{Fig:NewG2_DW_OmegaPi8} (and also Figure~\ref{Fig:NewSU3_DW_OmegaPi8}) precisely reconstruct the physically inequivalent portions\footnote{Upon submission of version 1 of this manuscript, we became aware of the preprint \cite{Tarrio:2013qga} where an exhaustive study of domain-walls and RG flows at $\omega \neq 0$ has been carried out in terms of the field variables $(z,\zeta_{12})$. Therein, the sets of flows in Figures~\ref{Fig:Old_DW_OmegaPi8} and \ref{Fig:NewG2_DW_OmegaPi8} have been shown to combine together and determine a unique cone of physically inequivalent flows at $\omega=\frac{\pi}{8}$ having the flows to the ${\mathcal{N}=1}\,\&\,\textrm{G}_{2}\,$ and ${\mathcal{N}=1}\,\&\,\textrm{G}^{*}_{2}\,$ points as boundaries.} (half of the left plot and one quarter of the right plot) of figure~$6$ in ref.~\cite{Tarrio:2013qga}. Remarkably, the very convenient parameterisation we used to perform the supergravity computations in the previous sections, is also adequate to capture all the types of flows at $\omega=\frac{\pi}{8}$.

In the second case, there are flows connecting the ${\mathcal{N}=8}\,\&\,\textrm{SO}(8)$ point at $\lambda_{0}=\lambda'_{0}=0$ to the \mbox{${\mathcal{N}=1\,\&\,\textrm{G}_{2}}$} and the genuine ${\mathcal{N}=1}\,\&\,\textrm{SU}(3)\,$ points located at
\beq
\begin{array}{lccccc}
{\mathcal{N}=1}\,\&\,\textrm{G}_{2} & \hspace{1mm} : \hspace{1mm} & \lambda_{0}=\lambda'_{0}=0.329  & , & \alpha_{0}=\phi_{0}=0.373 \pi  & , \\[1mm]
{\mathcal{N}=1}\,\&\,\textrm{SU}(3)  & \hspace{1mm} : \hspace{1mm} & \lambda_{0}=0.275 \,\, , \,\, \lambda'_{0}=0.573  & , & \alpha_{0}=\frac{3\pi}{4} \,\, , \,\, \phi_{0}= \frac{\pi}{4}  &  .
\end{array}
\eeq 
The corresponding values of the superpotential are $|W_{0}|_{\textrm{SO}(8)}={1}$, $|W_{0}|_{\textrm{G}_{2}}={1.151}$ and $|W_{0}|_{\textrm{SU}(3)}={1.316}$. When approaching the ${\mathcal{N}=1}\,\&\,{\textrm{SU}(3)}$ point, the mass spectrum and the eigenvalues of ${\Delta^{i}}_{j}$ are given by
\beq
\label{masses_Delta_SU3}
\begin{array}{lccccc}
{\mathcal{N}=1}\,\&\,\textrm{SU}(3) & \hspace{2mm} : \hspace{5mm} & m^2 L^2 &=&   6.449 \, , \, 6.449  \,\, , \,\, 1.551 \,\, ,\,\,  1.551 & , \\[1mm]
& &  \Delta &=& -1.449 \, , \,  -1.449 \,\, , \,\, 3.449 \,\, ,\,\,  3.449 &  .
\end{array}
\eeq 
Supersymmetry again requires the eigenvalues of ${\Delta^{i}}_{j}$ to come in pairs adding to $2$. This point then corresponds to two irrelevant operators of dimension $\,3-\Delta\,$ ($\Delta<0$) and two normalisable modes ($\Delta>\frac{3}{2}$). We have numerically determined the steepest descent trajectories and found a regular behaviour: they smoothly lie inside a (one parameter family) cone of flows from the  ${\mathcal{N}=8}\,\&\,\textrm{SO}(8)\,$ to the  ${\mathcal{N}=1}\,\&\,\textrm{SU}(3)\,$ point passing arbitrarily close to the  ${\mathcal{N}=1}\,\&\,\textrm{G}_{2}\,$ point. This is shown in Figure~\ref{Fig:NewSU3_DW_OmegaPi8}. As a final comment, the ${\mathcal{N}=1}\,\&\,\textrm{SU}(3)\,$ point turns out to be the one with the lowest energy at $\omega=\frac{\pi}{8}$.

\section{Summary $\&$ final remarks}
\label{sec:Final_remarks}

In this paper we have revised the SU(3)-invariant sector of the one-parameter family of SO(8) gauged supergravities discovered in ref.~\cite{Dall'Agata:2012bb}. Using the powerful framework of the embedding tensor, we performed a supergravity derivation of the scalar Lagrangian (section~\ref{sec:SU3_Lagrangian}), the fermion mass terms (section~\ref{sec:SU3_fermi_square_terms} $+$ appendix~\ref{App:fermi_masses}) and the $\mathcal{N}=2$ superpotential(s) (section~\ref{sec:superpotentials}) as a function of the electromagnetic phase $\omega$ and the six real scalars in the theory. 

The precise knowledge of the fermi mass terms allowed us to check the stability of a non-supersymmetric AdS critical point preserving $\textrm{SU}(3)$ symmetry which only exists for $\omega \neq 0$, hence being genuinely \textit{dyonic}. We find that this AdS solution is \textit{fully stable} under scalar fluctuations \textit{and} has a mass spectrum that is sensitive to the electromagnetic phase. As mentioned in the main text, this is the only example (up to our knowledge) of such a critical point in new maximal supergravity.

\begin{figure}[t!]
\begin{center}
\includegraphics[width=110mm]{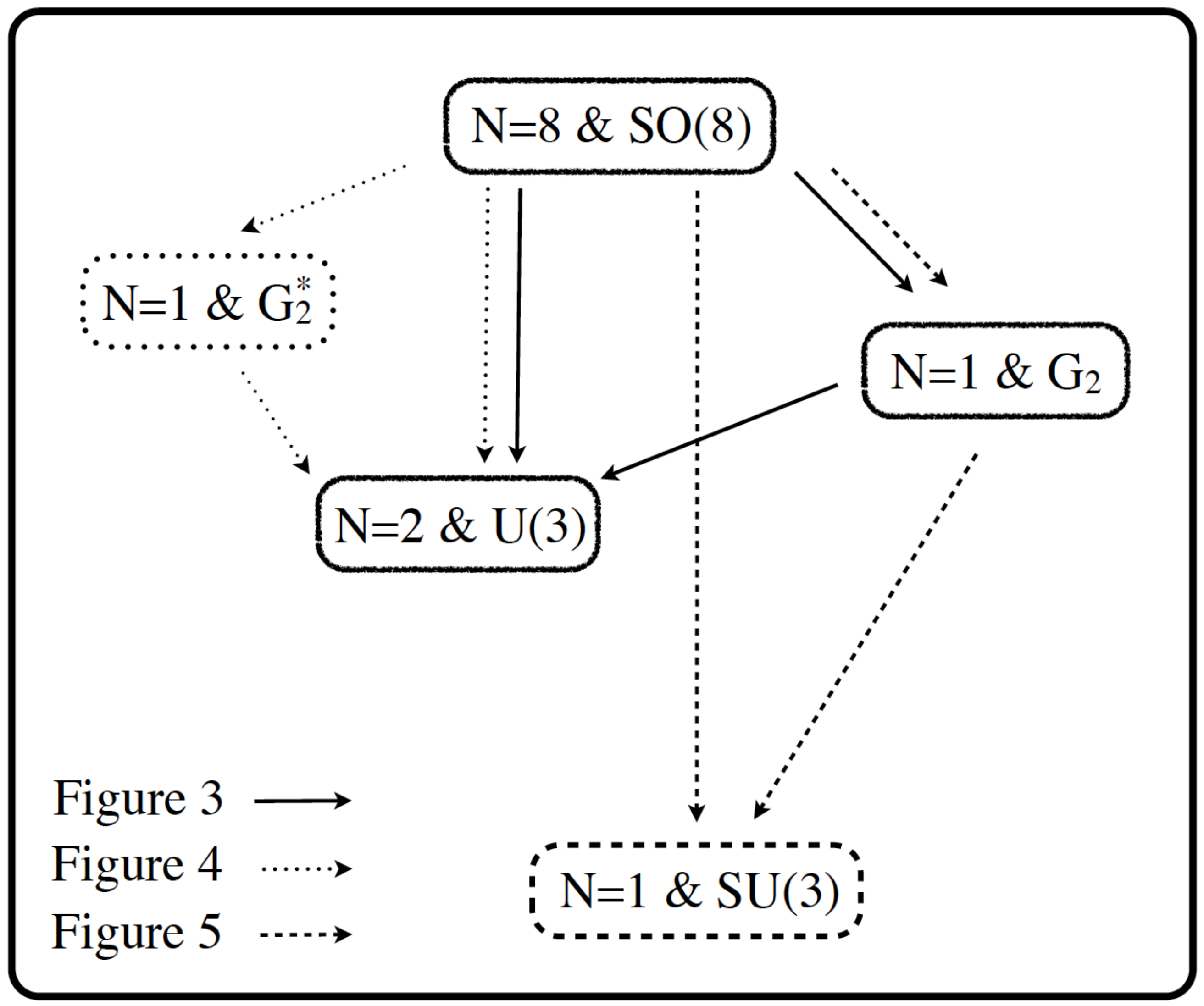}
\caption{{\it Flows between supersymmetric AdS points at $\omega=\frac{\pi}{8}$.}}
\label{Fig:Diagram} 
\end{center}
%\vspace{-5.2mm}
\end{figure}

In the second part of the paper, we presented some first results on BPS domain-walls for \mbox{$\omega \neq 0$}. Making use of the $\omega$-dependent superpotential(s) in (\ref{W1&What1}), we derived the first-order flow equations in (\ref{BPS-equations-compact}) and solved them numerically at $\omega=\frac{\pi}{8}$. In this way we obtained various flows between the (descending in energy) ${\mathcal{N}=8}\,\&\,\textrm{SO}(8)\,$, ${\mathcal{N}=1}\,\&\,\textrm{G}^{*}_{2}\,$, ${\mathcal{N}=1}\,\&\,\textrm{G}_{2}\,$, \mbox{${\mathcal{N}=2}\,\&\,\textrm{U}(3)$} and ${\mathcal{N}=1}\,\&\,\textrm{SU}(3)$ supersymmetric AdS points in Tables~\ref{Table:c=sqrt2-1_points} and \ref{Table:c=sqrt2-1_new_points} (see Figure~\ref{Fig:Diagram}). Some of them have a purely electric counterpart and behave in a similar way, \textit{e.g}. steepest descents smoothly lie inside the bounding cone. The others will not have such a smooth behaviour as they flow towards or pass nearby an AdS point which simply does not exist at $\omega=0$. In these cases, it would be very interesting to explore how the bounding cones blow up when taking the limit $\omega \rightarrow 0$ in which the ${\mathcal{N}=1}\,\&\,\textrm{G}^{*}_{2}\,$ and ${\mathcal{N}=1}\,\&\,\textrm{SU}(3)\,$ points run off to infinity in field space \cite{Borghese:2012zs}.  A dedicated study of domain-walls in \textit{dyonic} gauged supergravities will be presented somewhere else.

We would like to finish by commenting on potential applications of our results and also future directions. The first one concerns the search for a reduction Ansatz of 11d supergravity that could accommodate the electromagnetic phase $\omega$. To this end, if it is at all possible, the knowledge of the $T$-tensor and the fermion mass terms could play a central role \cite{Nicolai:2011cy}. We have derived these quantities as a function of the phase $\omega$ and the scalars in the $\textrm{SU}(3)$-invariant sector. This sector of the theory already encompasses many of the AdS points for which an 11d lifting could be figured out in the case of an electric $\textrm{SO}(8)$ gauging ($\omega=0$). For this reason, we believe that the $\omega$-dependent expressions obtained here might help in getting some insights in this direction. A second remark concerns the conjectured three-dimensional RG flows that the BPS domain-walls at $\omega = \frac{\pi}{8}$ would be dual to. In the case of $\omega=0$, these were connected to deformations of the BLG theory of M$2$-branes by a mass term of the form \cite{Ahn:2008ya,Ahn:2008gda,Bobev:2009ms}
\beq
\Delta W_{\textrm{BLG}} = \tfrac{1}{2} \, m_{1} \, \Phi^2_{1} +  \tfrac{1}{2} \, m_{\hat{1}} \, \Phi^2_{\hat{1}} \ ,
\eeq
and the bounding cone for the steepest descents in Figure~\ref{Fig:Old_DW_Omega0} was related to the $(m_{1},m_{\hat{1}})$ mass parameters. Therefore, a possible generalisation to $\omega \neq 0$ again demands the role of the electromagnetic phase to be better understood in the context of 11d supergravity. That goes beyond the scope of this work. Here, our aim was to construct flows between supersymmetric AdS points in new maximal supergravity. Nevertheless, the types of flows that we obtained for a \textit{dyonic} $\textrm{SO}(8)$ gauging could help in this task. We hope to come back to these issues in the future.

%%%%%%%%%%%%%%%%%%%%%%%%%%%%%%%%%%%%
%
% Acknowledgments
%
%%%%%%%%%%%%%%%%%%%%%%%%%%%%%%%%%%%%

%\vspace{-3mm}

\section*{Acknowledgments}
%\vspace{-2mm}

We want to thank J.~Armas and A.~Rothkopf for stimulating conversations. We especially thank J.P.~Derendinger for discussions and very useful explanations on R-symmetries and supermultiplets. And also P.~Nevado for providing helpful tips to solve differential equations numerically. The work of AG is supported by the Swiss National Science \mbox{Foundation}.

%%%%%%%%%%%%%%%%%%%%%%%%%%%%%%%%%%%%
%
% Appendices
%
%%%%%%%%%%%%%%%%%%%%%%%%%%%%%%%%%%%%

\appendix

\newpage

\section{Unitary gauge $\&$ $\textrm{E}_{7(7)}/\textrm{SU}(8)$ parameterisation}
\label{App:E7_parameterisaton}

The $\mathcal{N}=8$ supergravity multiplet in four dimensions contains $70$ real scalars which parameterise an element $\mathcal{V}$ of the coset space $\frac{\textrm{E}_{7(7)}}{\textrm{SU}(8)}\,$. The SU(8) in the denominator represents the maximal compact subgroup, so, in the unitary gauge, the physical scalars are associated to the non-compact generators of $\textrm{E}_{7(7)}$. In order to build the $56 \times 56$ coset representative ${\mathcal{V}_{\underline{\mathbb{M}}}}^{\underline{\mathbb{N}}} \in \frac{\textrm{E}_{7(7)}}{\textrm{SU}(8)}$ with both indices in the SU(8) basis, we will make extensive use of the $\Gamma$-matrices of $\textrm{SO}(8)\subset \textrm{SU}(8)$ we discuss now.

Majorana SO(8) spinors will be defined with an index down $\chi_{\mu}$. For the $\Gamma$-matrices and the charge conjugation matrix $\mathcal{C}$ we adopt the conventions in ref.~\cite{GSW_book}
\beq
{[\Gamma^{a}]_{\mu}}^{\nu} = \left(
\begin{array}{cc}
0  &  \left[\gamma^{a} \right]_{\alpha \dot{\beta}} \\
\left[ \bar{\gamma}^{a} \right]^{\dot{\alpha} \beta}  & 0\\  
\end{array} 
\right)
\hspace{10mm} \textrm{and} \hspace{10mm }
\mathcal{C}_{\mu\nu} = \left(
\begin{array}{cc}
\mathbb{I}_{\alpha \beta}  &  0 \\
0  & \mathbb{I}_{\dot{\alpha} \dot{\beta}} \\  
\end{array} 
\right) \ ,
\eeq
where $\,a=1,...,8\,$ is the vector index transforming in the $\bold{8}_{v}$, $\,\mu=1,...,16\,$ is a Majorana spinorial index and ${\,\alpha \,,\, \dot{\alpha} = 1,...,8\,}$ are left- and right-handed Majorana-Weyl spinorial indices transforming in the $\bold{8}_{s}$ and the $\bold{8}_{c}$, respectively. The splitting of the Majorana index is then of the form $_{\mu} =\, _{\alpha} \oplus\, ^{\dot{\alpha}}$. The matrices $\mathcal{C}=\mathcal{C}_{\mu\nu}$ and $\mathcal{C}^{-1}=\mathcal{C}^{\mu\nu}$ can be used to lower and rise indices resulting in $\Gamma$-($p$)forms
\beq
\label{Gamma(p)-form}
[\Gamma^{a_{1}...a_{p}}]_{\mu\nu}=\Gamma^{[a_{1}} \cdots \Gamma^{a_{p}]} \, \mathcal{C} \ ,
\eeq
with definite symmetry properties. The $\gamma^{a}$ building blocks are the $8 \times 8$ matrices
\beq
\begin{array}{cclccclc}
\gamma^{1} & = & i \sigma_{2} \otimes   i \sigma_{2} \otimes  i \sigma_{2} & \hspace{5mm} , \hspace{5mm} & \gamma^{5} & = & \sigma_{3} \otimes   i \sigma_{2} \otimes  \mathbb{I}_{2} & ,\\
\gamma^{2} & = & \mathbb{I}_{2} \otimes   \sigma_{1} \otimes  i \sigma_{2} & \hspace{5mm} , \hspace{5mm} & \gamma^{6} & = & i \sigma_{2} \otimes   \mathbb{I}_{2} \otimes  \sigma_{1} & ,\\
\gamma^{3} & = & \mathbb{I}_{2} \otimes   \sigma_{3} \otimes  i \sigma_{2} & \hspace{5mm} , \hspace{5mm} & \gamma^{7} & = & i \sigma_{2} \otimes  \mathbb{I}_{2} \otimes  \sigma_{3} & ,\\
\gamma^{4} & = & \sigma_{1} \otimes   i \sigma_{2} \otimes  \mathbb{I}_{2} & \hspace{5mm} , \hspace{5mm} & \gamma^{8} & = & \mathbb{I}_{2} \otimes  \mathbb{I}_{2} \otimes \mathbb{I}_{2} & ,
\end{array}
\eeq
built from the standard $\sigma_{1,2,3}$ Pauli matrices. As a result one finds $[\bar{\gamma}^{a}]=[\gamma^{a}]^{t}$, as well as the Clifford algebra
\beq
\gamma^{a} \, \bar{\gamma}^{b} + \gamma^{b} \, \bar{\gamma}^{a} = 2 \delta^{ab} \, \mathbb{I}_{8} 
\hspace{10mm} \textrm{ with } \hspace{10mm}
\delta^{ab}=\mathbb{I}_{8} \ ,
\eeq
so the matrices $\delta^{ab}$ and $\delta_{ab}$ can then be used to rise and lower indices in the $\textbf{8}_{v}$.

Two relevant $\Gamma$-($p$)forms in (\ref{Gamma(p)-form}) are those for $p=2,4$. Out of these, we can extract the pieces
\beq
\begin{array}{ccccl}
p=2 : \hspace{10mm} &  [\gamma^{ab}]_{\alpha\beta} & , & [\bar{\gamma}^{ab}]_{\dot{\alpha}\dot{\beta}}  & \hspace{5mm} \textrm{(antisymmetric)} \\[1mm]
p=4 : \hspace{10mm} &  [\gamma^{abcd}]_{\alpha\beta} & , & [\bar{\gamma}^{abcd}]_{\dot{\alpha}\dot{\beta}} & \hspace{5mm} \textrm{(symmetric)} 
\end{array}
\eeq
The former ($p=2$) are used to build the change of basis in (\ref{Rot_U}) after the index identification $a \leftrightarrow I$ and $\alpha \leftrightarrow A$. This translates into:  $\,[\gamma^{ab}]_{\alpha\beta} \leftrightarrow [\gamma^{IJ}]_{AB}$, etc. The latter ($p=4$) satisfy the (anti)self-duality conditions
\beq
\label{gamma4}
 [\gamma_{abcd}]_{\alpha\beta} = \frac{1}{4!} \,  \epsilon_{abcdefgh} \, [\gamma^{efgh}]_{\alpha\beta}
\hspace{8mm} \textrm{ and } \hspace{8mm}
  [\bar{\gamma}_{abcd}]_{\dot{\alpha}\dot{\beta}} = - \frac{1}{4!} \,  \epsilon_{abcdefgh} \, [\bar{\gamma}^{efgh}]_{\dot{\alpha}\dot{\beta}}
\eeq
and are used to split the $70$ real scalars in the theory into self-dual (SD) and anti-self-dual (ASD) ones transforming in the $\bold{35}_{s}$ and $\bold{35}_{c}$ of SO(8), respectively.  Then, the self-duality condition (\ref{SD_condition}) for complex scalars transforming in the $\textbf{70}$ of SU(8) is automatically fulfilled by the combination $\textbf{70}=\textbf{35}_{s} + i \,\textbf{35}_{c}$.

The next step is to build the $56 \times 56$ generators of $\textrm{E}_{7(7)}$ in the SU(8) basis, identify the non-compact ones and exponentiate them to build the coset representative ${\mathcal{V}_{\underline{\mathbb{M}}}}^{\underline{\mathbb{N}}}$. Taking (\ref{gamma4}) as the starting point, a systematic construction of the $\textrm{E}_{7(7)}$ generators in the SU(8) basis is explained in detail in the very useful appendix~A of ref.\cite{Fischbacher:2009cj}. Following the prescription there, the $\textrm{E}_{7(7)}$ generators $\,t_{\mathsf{A}}\,$ with $\mathsf{A}=1,...,133$ being the adjoint index, are organised as
\beq
t_{\mathsf{A}} \,\, \rightarrow \,\, \underbrace{\underbrace{t_{\mathsf{A}=1,...,35}}_{\textbf{35}_{s}} \,\, \oplus \,\, \underbrace{t_{\mathsf{A}=36,...,70}}_{\textbf{35}_{c}}}_{70 \,\, \textrm{non-compact}}  \,\, \oplus \,\, \underbrace{ \underbrace{t_{\mathsf{A}=71,...,105}}_{\textbf{35}_{v}} \,\, \oplus \,\, \underbrace{t_{\mathsf{A}=106,...,133}}_{\textbf{28} \, \rightarrow \, \textrm{SO}(8)} }_{63 \,\, \textrm{compact} \, \rightarrow \, \textrm{SU}(8)} \ ,
\eeq
so that the physical scalars of the theory will be associated with the first $70$ generators. The coset representative is then explicitly built out of the generators as
\beq
\label{Vielbein}
{\mathcal{V}_{\underline{\mathbb{M}}}}^{\underline{\mathbb{N}}} = \textrm{Exp}\left[ \, \sum_{m=1}^{35} \varphi^{(s)}_{m} \, {[t_{m}]_{\underline{\mathbb{M}}}}^{\underline{\mathbb{N}}} \, + \, \sum_{m=1}^{35} \varphi^{(c)}_{m} \, {[t_{35+m}]_{\underline{\mathbb{M}}}}^{\underline{\mathbb{N}}} \,\right] \ ,
\eeq
where $\,\varphi^{(s)}_{m=1,...,35}\,$ and $\,\varphi^{(c)}_{m=1,...,35}\,$ account for all the real scalars of maximal supergravity. The $\textrm{SU}(8)$ self-dual four-form in (\ref{SD_condition}) is constructed as\footnote{We use normalised $\delta^{\alpha \beta \gamma \delta}_{IJKL}$ Kronecker symbols with weight one such that 
\beq
\Sigma_{1234} = \sum_{m=1}^{35} \, \left( \varphi^{(s)}_{m} \, [S^{m}]_{1234} \,+\, i \, \varphi^{(c)}_{m} \, [C^{m}]_{1234}  \right) \ ,
\eeq
and similarly for the rest of components of $\Sigma_{IJKL}$.}
\beq
\label{Sigma_expansion}
\Sigma_{IJKL} =  \sum_{m=1}^{35} \Big( \varphi^{(s)}_{m} \, [S^{m}]_{abcd}    \, + \, i \, \varphi^{(c)}_{m} \, [C^{m}]_{abcd} \Big) \, \delta^{abcd}_{IJKL} \ ,
\eeq
where the expression for the tensors $\,[S^{m}]_{abcd}\,$ and $\,[C^{m}]_{abcd}\,$ in terms of (\ref{gamma4}) can again be found in ref.\cite{Fischbacher:2009cj}.

With all the above ingredients, the prescription to build the \textit{mixed} coset representative $\,{\mathcal{V}_{\mathbb{M}}}^{\underline{\mathbb{N}}}\,$ entering the scalar matrix (\ref{scalar_matrix}) for a given $G_{0}$-invariant sector of maximal supergravity is as follows:
\begin{itemize}

\item[$1)$] The precise embedding of $G_{0}$ inside the R-symmetry group SU(8) specifies the set of $G_{0}$-invariant four-forms and therefore the set of components in $\,\Sigma_{IJKL}\,$ which are compatible with the residual symmetry. 

\item[$2)$] After identifying the $G_{0}$-invariant components inside $\,\Sigma_{IJKL}\,$, it is immediate to read off which fields $\varphi_{m}^{(s)}$ and $\varphi_{m}^{(c)}$ are activated in the expansion (\ref{Sigma_expansion}) and plug them into (\ref{Vielbein}) to obtain the coset representative $\,{\mathcal{V}_{\underline{\mathbb{M}}}}^{\underline{\mathbb{N}}}\,$ in the SU(8) basis.

\item[$3)$] Finally we obtain the \textit{mixed} $\textrm{E}_{7(7)}/\textrm{SU}(8)$ coset representative ${\mathcal{V}_{\mathbb{M}}}^{\underline{\mathbb{N}}}={[U^{-1}]_{\mathbb{M}}}^{\underline{\mathbb{P}}}\, {\mathcal{V}_{\underline{\mathbb{P}}}}^{\underline{\mathbb{N}}}$ by applying the (inverse) change of basis in (\ref{Rot_U}). This is the vielbein we need in order to obtain the scalar matrix $\mathcal{M}_{\mathbb{MN}}$ in (\ref{scalar_matrix}).

\end{itemize}

\subsubsection*{Example: $\textrm{SU}(3)$-invariant sector}

Let us work out explicitly the case of the $\textrm{SU}(3)$-invariant sector that we analyse in this paper. The expansion (\ref{Sigma-form}) singles out the set of $\textrm{SU}(3)$-invariant forms in (\ref{real-forms}) and (\ref{complex-forms}). Matching them to the expression in (\ref{Sigma_expansion}) picks out the scalars
\beq
\begin{array}{c}
\varphi^{(s)}_{6} =\frac{\varphi^{(s)}_{4}}{2} =\varphi^{(s)}_{2} \,\,\, , \,\,\, \varphi^{(s)}_{3}=2\, \varphi^{(s)}_{2}-\varphi^{(s)}_{1} \,\, \,,\, \,\, \varphi^{(s)}_{5}=\varphi^{(s)}_{2}-\varphi^{(s)}_{1} \,\, \,,\, \,\, \varphi^{(s)}_{7}=\varphi^{(s)}_{1} + \varphi^{(s)}_{2} \\[1mm]
\varphi^{(s)}_{15}=-\varphi^{(s)}_{2} + 2\, \varphi^{(s)}_{1}\,\, \,\,\,,\,\,\, \,\, \varphi^{(s)}_{32}=-\varphi^{(s)}_{2}- 2\, \varphi^{(s)}_{1}   \,\, \,\,\,,\,\,\, \,\, \varphi^{(s)}_{17}=\varphi^{(s)}_{20}=-\varphi^{(s)}_{22}=-\varphi^{(s)}_{25} \ ,
\end{array}
\eeq
as well as
\beq
\begin{array}{c}
\varphi^{(c)}_{6} =\frac{\varphi^{(c)}_{4}}{2} =\varphi^{(c)}_{2} \,\,\, , \,\,\, \varphi^{(c)}_{1}= \varphi^{(c)}_{2}+\varphi^{(c)}_{7} \,\, \,,\, \,\, \varphi^{(c)}_{3}=\varphi^{(c)}_{2}-\varphi^{(c)}_{7} \,\, \,,\, \,\, \varphi^{(c)}_{5}=2\,\varphi^{(c)}_{2} - \varphi^{(c)}_{7} \\[1mm]
\varphi^{(c)}_{10}=\varphi^{(c)}_{2} + 2\, \varphi^{(c)}_{7}\,\, \,\,\,\,\,,\,\,\,\,\, \,\, \varphi^{(c)}_{33}=\varphi^{(c)}_{2}- 2\, \varphi^{(c)}_{7}   \,\, \,\,\,\,\,,\,\,\,\,\, \,\, \varphi^{(c)}_{12}=\varphi^{(c)}_{27}=-\varphi^{(c)}_{13}=-\varphi^{(c)}_{28} \ .
\end{array}
\eeq
In terms of the scalars in (\ref{Fields_of_V}), the independent fields are given by
\beq
\begin{array}{cccccccc}
\varphi^{(s)}_{1} &=& -\frac{1}{4} \lambda \cos(\alpha)  &  ,  & \varphi^{(s)}_{2} &=&  -\frac{1}{2} \lambda' \cos(\phi) \cos(\theta+\psi) & , \\[1mm]
\varphi^{(c)}_{7} &=&  -\frac{1}{4} \lambda \sin(\alpha) &  ,  & \varphi^{(c)}_{2} &=& -\frac{1}{2} \lambda' \sin(\phi) \cos(\theta-\psi)  & , \\
\end{array}
\eeq
and, after multiplication by the corresponding $\textrm{E}_{7(7)}$ generators, they completely determine the coset representative ${\mathcal{V}_{\underline{\mathbb{M}}}}^{\underline{\mathbb{N}}}(\lambda, \alpha,\lambda',\phi,\theta,\psi)$ in (\ref{Vielbein}). The ultimate \textit{mixed} vielbein ${\mathcal{V}_{\mathbb{M}}}^{\underline{\mathbb{N}}}(\lambda, \alpha,\lambda',\phi,\theta,\psi)$ entering (\ref{scalar_matrix}) is then obtained by acting with ${[U^{-1}]_{\mathbb{M}}}^{\underline{\mathbb{P}}}$.

\section{Gravitino-dilatino mass terms}
\label{App:fermi_masses}

In this appendix we present the explicit form of the ${\mathcal{A}_{I}}^{JKL}$ tensor corresponding to the gravitino-dilatino mass terms in (\ref{fermi_couplings}).
\vspace{4mm}

\noindent $\circ$ \textit{Couplings involving the $\,\psi^{1}_{\mu}\,$ gravitino}

\vspace{2mm}

\noindent There are three of these couplings in (\ref{fermi_couplings}). Their expressions are
\beq
\label{A2_Psi_1-Terms}
\begin{array}{ccl}
\\[-5mm]
{\mathcal{A}_{+1}}^{abc} & = & \tfrac{e^{-3 i (\alpha +\phi )}}{2}  \sinh (\lambda ) \sinh (2 \lambda' ) \Big[ 3 \,  e^{2 i (\alpha +\phi )} \cosh ^2(\lambda ) \cosh (2 \lambda' ) + \sinh ^2(\lambda )   j_{1}(\lambda',\phi) \Big]  \\[3mm]
{\mathcal{A}_{-1}}^{abc} & = & \,\,\,\, \frac{e^{-3 i \phi }}{2}  \cosh (\lambda ) \sinh (2 \lambda' ) \Big[ 3 \, e^{-2 i (\alpha -\phi )} \sinh ^2(\lambda ) \cosh (2 \lambda' ) + \cosh ^2(\lambda ) j_{1}(\lambda',\phi) \Big] \\[5mm]
{\mathcal{A}_{+1}}^{\hat{1} a\hat{a}} & = & -\frac{e^{-2 i (\alpha +2 \phi )}}{4}  \cosh (\lambda ) \Big[ e^{2 i (\phi+\alpha )} h_{1}(\lambda) \sinh ^2(2 \lambda' )+4 \, \sinh ^2(\lambda ) f_{2}(\lambda',\phi)  \Big]  \\[3mm]
{\mathcal{A}_{-1}}^{\hat{1} a\hat{a}} & = & \,\,\, \,-\frac{e^{i (\alpha -4 \phi )}}{4}  \sinh (\lambda ) \, \Big[  e^{2 i (\phi-\alpha) } h_{2}(\lambda) \sinh ^2(2 \lambda' ) +4 \,\cosh ^2(\lambda ) f_{2}(\lambda',\phi) \Big] 
\end{array}
\eeq
together with the additional ${\mathcal{A}_{\pm1}}^{a\hat{b}\hat{c}}=-{\mathcal{A}_{\pm1}}^{abc}$. The scalar-dependent functions entering the above couplings read 
\beq
\begin{array}{rclcrclc}
f_{2}(\lambda',\phi) & = & \sinh ^4(\lambda' )+e^{4 i \phi } \cosh ^4(\lambda' ) & , & g_{2}(\lambda') & = & \cosh (4 \lambda' )+3  & , \\
j_{1}(\lambda',\phi) & = & \sinh ^2(\lambda' )+e^{4 i \phi } \cosh ^2(\lambda' ) & , & h_{1}(\lambda) & = & 3 \cosh (2 \lambda )-1 & , \\
j_{2}(\lambda',\phi) & = & \cosh ^2(\lambda' )+e^{4 i \phi } \sinh ^2(\lambda' ) & , & h_{2}(\lambda) & = & 3 \cosh (2 \lambda )+1  & ,
\end{array}
\eeq
and are introduced to reduce the size of the expressions.
\vspace{4mm}

\noindent $\circ$ \textit{Couplings involving the $\,\psi^{\hat{1}}_{\mu}\,$ gravitini}
\vspace{2mm}

\noindent The situation for these couplings is analogous to the case of the $\,\psi^{1}_{\mu}\,$ gravitino discussed before. The set of couplings consists of
\beq
\label{A2_Psi_1h-Terms}
\begin{array}{ccl}
{\mathcal{A}_{+\hat{1}}}^{\hat{a}\hat{b}\hat{c}} & = & \frac{e^{-i (3 \alpha +\phi )}}{2}  \sinh (\lambda ) \sinh (2 \lambda' ) \Big[ 3 \, e^{2 i (\alpha +\phi )} \cosh ^2(\lambda ) \cosh (2 \lambda' ) + \sinh ^2(\lambda )   j_{2}(\lambda',\phi) \Big] \\[3mm]  
{\mathcal{A}_{-\hat{1}}}^{\hat{a}\hat{b}\hat{c}} & = & \,\,\,\,\,\,  \frac{e^{-i \phi }}{2}  \cosh (\lambda ) \sinh (2 \lambda' ) \Big[ 3 \, e^{-2 i (\alpha -\phi )} \sinh ^2(\lambda ) \cosh (2 \lambda' ) + \cosh ^2(\lambda )  j_{2}(\lambda',\phi) \Big]\\[5mm]
{\mathcal{A}_{+\hat{1}}}^{1 a\hat{a}} & = & \frac{e^{-2 i \alpha }}{4}  \cosh (\lambda ) \Big[ e^{2 i (\alpha +\phi )} h_{1}(\lambda) \sinh ^2(2 \lambda' )+4 \,  \sinh ^2(\lambda ) f_{1}(\lambda',\phi)  \Big] \\[3mm] 
{\mathcal{A}_{-\hat{1}}}^{1 a\hat{a}} & = &\,\,\,\, \frac{e^{i \alpha }}{4}  \, \sinh (\lambda )  \Big[  e^{2 i (\phi - \alpha)   } h_{2}(\lambda) \sinh ^2(2 \lambda' ) + 4 \,  \cosh ^2(\lambda ) f_{1}(\lambda',\phi) \Big] 
\end{array}
\eeq
together with  ${\mathcal{A}_{\pm\hat{1}}}^{ab\hat{c}}=-{\mathcal{A}_{\pm\hat{1}}}^{\hat{a}\hat{b}\hat{c}}$. A quick comparison between (\ref{A2_Psi_1-Terms}) and (\ref{A2_Psi_1h-Terms}) makes the similarities between the two SU(3)-singlet gravitini manifest.

\newpage

\noindent $\circ$ \textit{Couplings involving the $\,\psi^{a}_{\mu}\,$ gravitini}
\vspace{2mm}

\noindent There are five couplings between these gravitini and the set of dilatini in (\ref{fermi_couplings}). These are given by
\beq
\label{A2_Psi_a-Terms}
\begin{array}{ccl}
\\[-5mm]
{\mathcal{A}_{+a}}^{1\hat{b}\hat{c}} & = & \frac{e^{-i (\alpha +\phi )}}{8}  \sinh (2 \lambda' )  \Big[  4 \sinh (\lambda ) \cosh ^2(\lambda ) j_{2}(\lambda',\phi) -e^{2 i (\alpha +\phi )} r_{1}(\lambda) \cosh (2 \lambda' ) \Big]  \\[3mm]
{\mathcal{A}_{-a}}^{1\hat{b}\hat{c}} & = &    \frac{e^{i (2 \alpha - \phi) }}{8}  \sinh (2 \lambda' )  \Big[  4 \,  \sinh ^2(\lambda ) \cosh (\lambda ) j_{2}(\lambda',\phi) + e^{2 i (\phi-\alpha) } r_{2}(\lambda) \cosh (2 \lambda' ) \Big] \\[5mm]
{\mathcal{A}_{+a}}^{\hat{a} 1\hat{1}} & = &\,\,\,\,\,\,\,\,\,  -\frac{1}{4}  \cosh (\lambda ) \Big[  2 \cosh^2(\lambda ) \sinh ^2(2 \lambda' ) \cos(2 \phi) + e^{2 i \alpha} \sinh ^2(\lambda ) g_{1}(\lambda') \Big]  \\[3mm]
{\mathcal{A}_{-a}}^{\hat{a} 1\hat{1}} & = & -\frac{e^{3 i \alpha }}{4}  \sinh (\lambda ) \Big[  2 \, \sinh ^2(\lambda ) \sinh ^2(2 \lambda' ) \cos (2 \phi)  +  e^{-2 i \alpha }\cosh ^2(\lambda )  g_{1}(\lambda') \Big]  \\[5mm]
{\mathcal{A}_{+a}}^{\hat{a} b\hat{b}} & = & \,\,\, - \frac{1}{8} \cosh (\lambda ) \Big[  4 \, e^{-2 i \alpha } \sinh ^2(\lambda ) \sinh ^2(2 \lambda' ) \cos (2 \phi )  +  \cosh (2 \lambda ) g_{1}(\lambda') - g_{2}(\lambda') \Big]  \\[3mm]
{\mathcal{A}_{-a}}^{\hat{a} b\hat{b}} & = & -\frac{e^{-i \alpha }}{8}  \sinh (\lambda ) \Big[  4 \, e^{2 i \alpha } \cosh ^2(\lambda ) \sinh ^2(2 \lambda' ) \cos (2 \phi )+\cosh (2 \lambda ) g_{1}(\lambda')  +g_{2}(\lambda')  \Big]  \\[5mm]
{\mathcal{A}_{+a}}^{\hat{1} b \hat{c}} & = & \frac{e^{-i (\alpha +3 \phi )}}{4}  \sinh (\lambda ) \sinh (2 \lambda' ) \Big[ 2  \cosh ^2(\lambda ) j_{1}(\lambda',\phi) +  e^{2 i (\phi+\alpha )} h_{2}(\lambda) \cosh (2 \lambda' )\Big]   \\[3mm]
{\mathcal{A}_{-a}}^{\hat{1} b \hat{c}}& = &  \frac{e^{i( 2 \alpha - 3\phi) } }{4} \cosh (\lambda ) \sinh (2 \lambda' ) \Big[  2\,   \sinh^2 (\lambda ) j_{1}(\lambda',\phi) +  e^{2 i (\phi-\alpha) } h_{1}(\lambda) \cosh (2 \lambda' ) \Big] 
\end{array}
\eeq
as well as ${\mathcal{A}_{\pm a}}^{1 b c}=-{\mathcal{A}_{\pm a}}^{1 \hat{b}\hat{c}}$. The new  functions appearing in (\ref{A2_Psi_a-Terms}) are
\beq
\begin{array}{rclcrclc}
r_{1}(\lambda) & = & \sinh (\lambda )-3 \sinh (3 \lambda ) & , & r_{2}(\lambda) & = & \cosh (\lambda )+3 \cosh (3 \lambda )  & ,
\end{array}
\eeq
and complete the set of functions we will introduce to simplify the expressions. 

\vspace{4mm}

\noindent $\circ$ \textit{Couplings involving the $\,\psi^{\hat{a}}_{\mu}\,$ gravitini}
\vspace{2mm}

\noindent The couplings to these gravitini match those already found for their counterparts $\,\psi^{a}_{\mu}\,$. They are given by
\beq
\label{A2_Psi_ah-Terms}
\begin{array}{lclcccccc}
{\mathcal{A}_{\pm\hat{a}}}^{\hat{1} b c}={\mathcal{A}_{\pm a}}^{\hat{1} b \hat{c}} 
& , & 
{\mathcal{A}_{\pm\hat{a}}}^{a 1 \hat{1} }= -{\mathcal{A}_{\pm a}}^{\hat{a} 1 \hat{1} }
 & , & 
{\mathcal{A}_{\pm \hat{a}}}^{a b \hat{b}}=-{\mathcal{A}_{\pm a}}^{\hat{a} b \hat{b}} & ,
\\[1mm]
{\mathcal{A}_{\pm \hat{a}}}^{1 b \hat{c} }= {\mathcal{A}_{\pm a}}^{1 \hat{b} \hat{c} }
 & , & 
 {\mathcal{A}_{\pm\hat{a}}}^{\hat{1} \hat{b} \hat{c}}=- {\mathcal{A}_{\pm a}}^{\hat{1} b \hat{c}}
 & ,
\end{array}
\eeq
and complete the set of gravitino-dilatino couplings of the SU(3)-truncated theory.

\newpage

%%%%%%%%%%%%%%%%%%%%%%%%%%%%%%%%%%%%
%
% Bibliography
%
%%%%%%%%%%%%%%%%%%%%%%%%%%%%%%%%%%%%

\small

\bibliography{references}
\bibliographystyle{utphys}

\end{document}